\input harvmac 
\input amssym.def
\input amssym.tex
\parskip=4pt \baselineskip=12pt
\hfuzz=20pt 
\parindent 10pt 
\font\tfont=cmbx12 scaled\magstep1  

\global\secno=0

\font\male=cmr9 
 
\def\half{{\textstyle{1\over2}}}
\def\quater{{\textstyle{1\over4}}}
\def\ap{\approx} 

\font\fat=cmsy10 scaled\magstep5
 
\def\Bbullet{\raise-3pt\hbox{\fat\char"0F}}

\def\nt{\noindent} 
\def\nl{\hfil\break} 
\def\np{\vfil\eject} 
\def\re#1{\nt{\bf Remark #1:}~~} 
\def\pr#1{\nt{\bf Proposition #1:}~~}

\def\riga{-\kern-4pt - \kern-4pt -}
\def\driga{{\raise-2.2pt\hbox{=\kern-4pt = \kern-4pt =}
\kern-18.5pt\raise2.2pt\hbox{=\kern-4pt = \kern-4pt =}}
>}

\def\dia{~~$\diamondsuit$} 
\def\bsq{~~$\blacksquare$}
\def\bu{$\bullet$} 

  \def\cc{{\cal R}}
  \def\cf{{\cal F}} 
\def\cg{{\cal G}} \def\ch{{\cal H}} 
   
 \def\cn{{\cal N}}

\def\bbc{C\kern-6pt I} 
\def\bac{{C\kern-5.5pt I}} 

  \def\o{{\bar 0}} \def\I{{\bar 1}}

\def\a{\alpha}  \def\g{\gamma} \def\d{\delta} 

\def\eps{{\epsilon}}
  
\def\r{\rho} \def\om{\omega}

\def\D{\Delta} \def\L{\Lambda}

\def\hi{{\hat \imath}} \def\hj{{\hat \jmath}} 
 
\def\hk{{\hat k}} 
\def\hl{{\hat\ell}}\def\hm{{\hat m}} 
  
\def\tk{{\tilde k}} \def\tl{{\tilde\ell}}

\def\hH{{\hat H}} \def\tH{{\tilde H}}

\def\tnu{\tilde{\nu}} 
\def\ve{\varepsilon} 
\def\tve{\tilde{\ve}} 
\def\hnu{{\hat \nu}} \def\hve{{\hat \ve}}

\def\({\left(}
\def\){\right)}

\def\a{\alpha}         

\def\d{\delta}         
\def\g{\gamma}

\def\r{\rho}

\def\D{\Delta}         
\def\L{\Lambda}


\nref\LMRS{S. Lee, S. Minwalla, M. Rangamani and N. Seiberg, Adv.
Theor. Math. Phys. {\bf 2}, 697 (1998), hep-th/9806074.}

\nref\NS{A. Nelson and M.J. Strassler, Phys. Rev. {\bf D60}, 015004
(1999), hep-ph/9806346;\ J. High En. Phys. (9) U553 (2000),
hep-ph/0006251.}

\nref\Osb{H. Osborn, Ann. Phys. (NY) {\bf 272}, 243 (1999),
hep-th/9808041.}

\nref\Park{J.-H. Park, Nucl. Phys. {\bf B559}, 455 (1999),
hep-th/9903230.}

\nref\FGPW{D.Z. Freedman, S.S. Gubser, K. Pilch and N.P. Warner,
Adv. Theor. Math. Phys. {\bf 3}, 363 (1999), hep-th/9904017.}

\nref\GrKa{M. Gremm and A. Kapustin, 
J. High En. Phys. (7) U92 (1999), hep-th/9904050.} 

\nref\AGMOO{O. Aharony, S.S. Gubser, J. Maldacena, H. Ooguri and
Y. Oz, Phys. Rep. {\bf 323}, 184 (2000), hep-th/9905111.}
 
\nref\CDDF{A. Ceresole, G. Dall'Agata, R. D'Auria and S. Ferrara,
Phys. Rev. {\bf D61}, 066001 (2000), hep-th/9905226;\ 
Class. Quant. Grav. {\bf 17}, 1017 (2000), hep-th/9910066.}

\nref\DFMMR{E. D'Hoker, D.Z. Freedman, S.D. Mathur, A. Matusis
and L. Rastelli, in: Shifman, M.A. (ed.) 'The Many Faces of the
Superworld', p. 332, hep-th/9908160.}

\nref\BKRS{M. Bianchi, S. Kovacs, G. Rossi and Y.S. Stanev, 
J. High En. Phys. (8) U397 (1999), hep-th/9906188;\ 
Nucl. Phys. {\bf B584} (2000) 216, hep-th/0003203;\ 
J. High En. Phys. (5) U960 (2001), hep-th/0104016.} 

\nref\FeZa{S. Ferrara and A. Zaffaroni, to appear in the
proceedings of the Moshe Flato Conference, Dijon, 1999,
CERN-TH/99-255, hep-th/9908163.}

\nref\AFSZ{L. Andrianopoli, S. Ferrara, E. Sokatchev and B. Zupnik,
Adv. Theor. Math. Phys. {\bf 3}, 1149 (1999), hep-th/9912007.}

\nref\FS{S. Ferrara and E. Sokatchev, 
Lett. Math. Phys. {\bf 52}, 247 (2000), 
hep-th/9912168;\ 
J. High En. Phys. (5) U828 (2000), hep-th/0003051;\ 
Int. J. Mod. Phys. {\bf B14}, 2315 (2000), hep-th/0007058;\ 
J. Math. Phys. {\bf 42}, 3015 (2001), hep-th/0010117.}

\nref\Pelc{O. Pelc, J. High En. Phys. (3) U216 (2000),
hep-th/0001054.}

\nref\Fa{S. Ferrara, TMR Conference, Paris, 1-7 Sep 1999,
hep-th/0002141.} 

\nref\BaZu{F. Bastianelli and R. Zucchini, 
J. High En. Phys. (5) U1041 (2000), hep-th/0003230.} 

\nref\HeHo{P. Heslop and P.S. Howe, Class. Quant. Grav. {\bf 17},
3743 (2000), hep-th/0005135;\ TMR Conference, Paris, September
2000, hep-th/0009217;\ Phys. Lett. {\bf 516B}, 367 (2001),
hep-th/0106238;\ OPEs and 3-point correlators of protected
operators in N=4 SYM, hep-th/0107212.} 

\nref\FSa{S. Ferrara and E. Sokatchev, 
Int. J. Theor. Phys. {\bf 40}, 935 (2001), hep-th/0005151.} 

\nref\DLS{M.J. Duff, J.T. Liu and K.S. Stelle, 
J. Math. Phys. {\bf 42}, 3027 (2001) hep-th/0007120.}

\nref\LiSa{J.T. Liu and H. Sati, 
Nucl. Phys. {\bf B605}, 116 (2001), hep-th/0009184.}

\nref\KlSa{D. Klemm and W.A. Sabra, J. High En. Phys. (2)
U632 (2001), hep-th/0011016.} 

\nref\FrHL{D.Z. Freedman and P. Henry-Labordere, 
TMR Conference, Paris, 7-13 Sep 2000, hep-th/0011086.} 

\nref\MaWe{S.P. Martin and J.D. Wells, 
Phys Rev {\bf D64}, 036010 (2001), hep-ph/0011382.}

\nref\Arg{P.C. Argyres, Nucl. Phys. Proc. Sup. {\bf B101}, 175
(2001), hep-th/0102006.}
 
\nref\AEPS{G. Arutyunov, B. Eden, A.C. Petkou and E. Sokatchev,
Nucl. Phys. {\bf B620}, 380 (2002),  hep-th/0103230.}

\nref\AES{G. Arutyunov, B. Eden and E. Sokatchev, 
Nucl. Phys. {\bf B619}, 359 (2001), hep-th/0105254.}

\nref\EdSo{B. Eden and E. Sokatchev, Nucl. Phys. {\bf B618}, 259 (2001),
hep-th/0106249.}

\nref\PeSa{S. Penati and A. Santambrogio, 
Nucl. Phys. {\bf B614}, 367 (2001), hep-th/0107071.}

\nref\EFS{B. Eden, S. Ferrara and E. Sokatchev,
JHEP 0111:020 (2001), hep-th/0107084.}

\nref\KuSu{J. Kubo and D. Suematsu, Phys. Rev. {\bf D64}, 115014
(2001), hep-ph/0107133.} 

\nref\West{P. West, Nucl. Phys. Proc. Sup. {\bf B101}, 112 (2001).} 

\nref\Hes{P.J. Heslop, Class. Quant. Grav. {\bf 19}, 303 (2002), 
hep-th/0108235.}

\nref\Ryz{A.V. Ryzhov, JHEP 0111:046 (2001), hep-th/0109064.}

\nref\DHRy{E. D'Hoker and A.V. Ryzhov, Three-point functions of quarter BPS
operators in N=4 SYM, hep-th/0109065, UCLA/01/TEP/22.}

\nref\FSb{S. Ferrara and E. Sokatchev, to be published
in NJP Focus Issue: Supersymmetry in condensed matter and
high-energy physics, hep-th/0110174.} 


\lref\Kab{V.G. Kac, Adv. Math. {\bf 26}, 8-96 (1977);\ 
Comm. Math. Phys. {\bf 53}, 31-64 (1977);\ the second paper 
is an adaptation for physicists of the first paper.} 

\lref\Kc{V.G. Kac, Lect. Notes in Math. {\bf 676}
(Springer-Verlag, Berlin, 1978) pp. 597-626.}

\lref\DPm{V.K. Dobrev and V.B. Petkova, 
Lett. Math. Phys. {\bf 9}, 287 (1985).} 

\lref\DPf{V.K. Dobrev and V.B. Petkova, 
Fortschr. d. Phys. {\bf 35}, 537 (1987).} 

\lref\DPu{V.K. Dobrev and V.B. Petkova, 
Phys. Lett. {\bf 162B}, 127 (1985).}

\lref\DPp{V.K. Dobrev and V.B. Petkova, 
Proceedings, eds. A.O. Barut and H.D. Doebner, 
Lecture Notes in Physics, Vol. 261 (Springer-Verlag, Berlin, 1986) 
p. 291 
and p. 300.} 

\lref\FF{M. Flato and C. Fronsdal, Lett. Math. Phys. {\bf 8}, 159
(1984).} 

\lref\Sha{N.N. Shapovalov,
Funkts. Anal. Prilozh. {\bf 6} (4) 65 (1972);\ English translation:
Funkt. Anal. Appl. {\bf 6}, 307 (1972).}

\lref\Min{S. Minwalla, Adv. Theor. Math. Phys. {\bf 2}, 781
(1998).} 

\lref\DFLV{R. D'Auria, S. Ferrara, M.A. Lledo and V.S.
Varadarajan, Spinor algebras, hep-th/0010124, CERN-TH-2000-260,} 

\lref\GSB{M. Gunaydin and C. Saclioglu, Comm. Math. Phys. {\bf
87}, 159 (1982), ~Phys. Lett. {\bf 108B}, 180 (1982), (for
algebras); ~~I. Bars and M. Gunaydin, Comm. Math. Phys. {\bf 91},
31 (1983), ~M. Gunaydin, J. Math. Phys. {\bf 29}, 1275 (1988)
(for superalgebras).} 

\lref\GNWT{M.Gunaydin and P. van Nieuwenhuizen and N.P Warner, 
Nucl. Phys. {\bf B255}, 63 (1985), ~M. Gunaydin and S. Takemae, 
Nucl. Phys. {\bf B578}, 405 (2000).} 

\lref\HLS{R. Haag, J.T. Lopuszanski and M. Sohnius,
Nucl. Phys. {\bf B88}, 257 (1975).}

\lref\Nahm{W.~Nahm,
Nucl. Phys. {\bf B135}, 149 (1978).}

\lref\FeFr{S. Ferrara and C. Fronsdal, Conformal fields in higher
dimensions, hep-th/0006009.}


\hfill INRNE-TH-02-01 

\vskip 2truecm

\centerline{{\tfont Positive Energy Unitary Irreducible Representations}}

\vskip 2mm

\centerline{{\tfont of D=6 Conformal Supersymmetry}} 

\vskip 1.5cm

\centerline{{\bf V.K. Dobrev}} 

\vskip 0.5cm

\centerline{Institute of Nuclear Research and Nuclear Energy}
\centerline{Bulgarian Academy of Sciences}
\centerline{72 Tsarigradsko Chaussee, 1784 Sofia, Bulgaria}

\vskip 1.5cm

\centerline{{\bf Abstract}}

\midinsert\narrower\narrower{\male
We give a constructive classification of the positive energy
(lowest weight) unitary irreducible representations of 
the ~$D=6$~ superconformal algebras ~$osp(8^*/2N)$. Our results
confirm all but one of the conjectures of Minwalla (for N=1,2) 
on this classification. Our main tool is the explicit construction of
the norms of the states that has to be checked for positivity. 
We give also the reduction of the four exceptional UIRs. 
}\endinsert

\vskip 1.5cm

\newsec{Introduction} 

\nt 
Recently, superconformal field theories in various dimensions are
attracting more interest, in particular, due to their duality to
AdS supergravities, cf. [1-35] and references therein.
Particularly important are those for ~$D\leq 6$~ since in these
cases the relevant superconformal algebras satisfy \Nahm{} the
Haag-Lopuszanski-Sohnius theorem \HLS.  This makes the
classification of the UIRs of these superalgebras very important.
Until recently such classification was known only for the ~$D=4$~
superconformal algebras ~$su(2,2/N)$ \FF{} (for $N=1$),
\DPm,\DPu,\DPf,\DPp. Recently, the classification for ~$D=3$ 
(for even $N$), $D=5$, and $D=6$ (for $N=1,2$)~ was given in 
\Min{}, but some of the results were conjectural and there was not 
enough detail in order to check these conjectures. 
On the other hand the applications of $D=6$ unitary irreps 
require firmer theoretical basis. Among the many 
interesting applications we shall mention 
the analysis of OPEs and 1/2 BPS operators 
\FSa,\EFS,\FSb{}. In particular, it is important that 
some general properties of abstract superconformal field theories 
can be obtained by using the BPS nature of a certain class of 
superconformal primary operators and the model
independent nature of superconformal OPEs. 
In the classification of UIRs of superconformal algebras an 
important role is played by the representations with 
``quantized" conformal dimension since in the quantum field theory 
framework they correspond to operators with
``protected" scaling dimension and therefore imply ``non-renormalization
theorems" at the quantum level. 

Motivated by the above 
we decided to reexamine the list of UIRs of the ~$D=6$~ superconformal 
algebras in detail. More than that we treat the superalgebras 
~$osp(8^*/2N)$~ for arbitrary ~$N$. Thus, we give the final list 
of UIRs for ~$D=6$. With this we also confirm all but one of the 
conjectures of \Min{} for ~$N=1,2$. Our main tool is the
explicit construction of the norms. This, on the one hand,
enables us to prove the unitarity list, and, on the other hand, 
enables us to give explicitly the states of the irreps. 

The paper is organized as follows. In the Section 2 we
discuss in detail the lowest weight representations of the superalgebras
~$osp(8^*/2N)$. In particular, we define explicitly the norm
squared of the states that has to be checked for positivity. 
In Section 3 we state the main result (Theorem) on the lowest weight
(positive energy) UIRs and show explicitly the Proof of necessity. 
(After the Theorem we comment exactly on the results of \Min{} 
giving also the relation between our notations.) 
We also give the general
form of the norms which is enough for the Proof of sufficiency.
For part of the states (the fully factorizable ones) 
we give the norms explicitly, for the rest (the unfactorizable
ones) the formulae are very involved and in general only
recursive. These results are in the generic situation. In Section
4 we show the unitarity at the four exceptional points. We give
explicitly the states of zero norm (though not all for $N>1$),
which have to be decoupled for the unitary irrep. In Section 5
we discuss the ongoing research.

\newsec{Representations of D=6 conformal supersymmetry} 

\subsec{The setting}

\nt 
Our basic reference for Lie superalgebras is \Kab. 
The superconformal algebras in $D=6$ are ~$\cg ~=~ osp(8^*/2N)$. 
We label their physically relevant representations by the 
signature:
\eqn\sgn{\chi ~=~ [\, d\,;\, n_1\,,\, n_2\,,\, n_3\,; 
\,a_1\,,...,a_N\,]}
where ~$d$~ is the conformal weight, ~$n_1,n_2,n_3$~ are 
non-negative integers which are Dynkin labels of the finite-dimensional 
irreps of the $D=6$ Lorentz algebra ~$so(5,1)$, and ~$a_1,...,a_N$~ are 
non-negative integers which are Dynkin labels of the finite-dimensional 
irreps of the internal (or $R$) symmetry algebra ~$usp(2N)$.
The even subalgebra of ~$osp(8^*/2N)$~ is the algebra 
~$so^*(8) \oplus usp(2N)$, and ~$so^*(8)\cong so(6,2)$~ is the 
$D=6$ conformal algebra. 

Our aim is to give a constructive proof for the UIRs of ~$osp(8^*/2N)$~ 
following the methods used for the $D=4$ superconformal algebras 
~$su(2,2/N)$, cf. \DPu,\DPp,\DPf. The main tool is an adaptation 
of the Shapovalov form on the Verma modules ~$V^\chi$~ over the 
complexification ~$\cg^\bac ~=~ osp(8/2N)$~ of ~$\cg$. 

\subsec{Verma modules}

\nt 
To introduce Verma modules we use the standard 
triangular decomposition:
\eqn\trig{ \cg^\bac ~=~ \cg^+ \oplus \ch \oplus \cg^-}
where $\cg^+$, $\cg^-$, resp., are the subalgebras corresponding 
to the positive, negative, roots, resp., and $\ch$ denotes the 
Cartan subalgebra. 

We consider lowest weight Verma modules, 
so that ~$V^\L ~ \cong U(\cg^+) \otimes v_0\,$, 
 where ~$U(\cg^+)$~ is the universal enveloping algebra of $\cg^+$, 
and ~$v_0$~ is a lowest weight vector $v_0$ such that:
\eqn\low{\eqalign{ 
 Z \ v_0\ =&\ 0 \ , \quad Z\in \cg^- \cr 
 H \ v_0 \ =&\ \L(H)\ v_0 \ , \quad H\in \ch}}
Further, for simplicity we omit the sign
~$\otimes \,$, i.e., we write $p\,v_0\in V^\L$ with $p\in U(\cg^+)$. 

The lowest weight $\L$ is characterized by its values on the Cartan 
subalgebra ~$\ch$. In order to have ~$\L$~ corresponding to ~$\chi\,$, 
 one can choose a basis in ~$\ch$~ so that 
to obtain the entries in the signature $\chi$ by evaluating $\L$ 
on the basis elements of ~$\ch$. 

\subsec{Root systems} 

\nt 
In order to explain how the above is done 
we recall some facts about ~$osp(8/2N)$ 
(denoted $D(4,N)$ in \Kab). 
\foot{These initial facts can be given for ~$osp(2M/2N) = 
D(M,N)$~ in a very similar fashion.} 
Their root systems are given in terms of 
~$\eps_1\,\dots,\eps_{4}\,$, ~$\d_1\,\dots,\d_{N}\,$, 
~$(\eps_i,\eps_j) ~=~ \d_{ij}\,$, ~$(\d_\hi,\d_\hj) ~=~
-\d_{\hi\hj}\,$, ~$(\eps_i,\d_\hj) ~=~ 0$. 
The indices ~$i,j,...$~ will take values in the set ~$\{1,2,3,4\}$,
the indices ~$\hi,\hj,...$~ will take values in the set
~$\{1,...,N\}$. 
The even and odd roots systems are \Kab:
\eqn\dmn{
\D_\o ~~=~~ \{\pm \eps_i\pm\eps_j ~, ~~i< j ~, 
~~\pm\d_\hi\pm\d_\hj~, ~~\hi< \hj ~, 
~~\pm 2\d_\hi \} ~,
\qquad \D_\I ~~=~~ \{ \pm \eps_i\pm\d_\hj \}}
(we remind that the signs ~$\pm$~ are not correlated).\foot{The 
roots ~$\pm\eps_i\pm\eps_j$~ provide the root system of 
~$so(8;\bbc)$, 
the roots ~$\pm\d_i\pm\d_j$~ and ~$\pm 2\d_i$~ provide the root 
system of ~$sp(2N;\bbc)$.} 
We shall use the following simple root system \Kab: 
\eqna\ssdm
$$\eqalignno{ 
& \Pi  ~=~  \{\, \eps_{1}-\eps_2\,, \eps_{2}-\eps_3\,, \eps_{3}-\eps_4\, 
, \eps_{4}-\d_1 \, ,\,  \d_1-\d_2 \, ,\, 
 , \dots, \d_{N-1}- \d_{N} \, ,\, 2 \d_N\, \} \ , &\ssdm {a}}$$
or introducing standard notation for the simple roots:
$$\eqalignno{&\Pi ~=~  \{\,\a_1\,,...,\,\a_{4+N} \, \}&\ssdm {b} \cr 
&\a_j ~=~ \eps_{j}-\eps_{j+1}\ , \quad 
j=1,2,3 \cr 
&\a_4 ~=~ \eps_{4}-\d_1 \cr
&\a_{4+\hj} ~=~ \d_{\hj}-\d_{\hj+1}\ , \quad 
\hj=1,...,N-1 \cr
& \a_{4+N} ~=~ 2\d_{N}}$$
The root ~$\a_4 = \eps_{4}-\d_1$~ is odd, the other simple roots 
are even. For future use we need also 
the positive root system corresponding to ~$\Pi$~: 
\eqn\psdm{ \D_\o^+ ~~=~~ \{ \eps_i\pm\eps_j ~, ~~i< j ~, 
~~\d_\hi\pm\d_\hj~, ~~\hi< \hj ~, ~~ 2\d_\hi \} ~,
\qquad \D_\I^+ ~~=~~ \{ \eps_i\pm\d_\hj \}} 

\subsec{Basis of the Cartan subalgebra} 

\nt 
Let us denote by ~$H_A$~ the generators of the Cartan 
subalgebra, $A=1,...,4+N$. There is a standard choice 
for these generators \Kab{}. Namely, to every even simple root ~$\a_A$~ 
we choose a generator ~$H_A$~ so that the 
following equality is valid for arbitrary ~$\mu\in \ch^*$~:
\eqn\corr{ \mu (H_A) ~=~ (\mu , \a_A^\vee ) \ , \quad A\neq 4,}
where ~$\a_A^\vee ~\equiv~ 2\a_A / (\a_A,\a_A)$. 
Because these ~$H_A$~ correspond to the simple even roots, 
which define the Dynkin labelling, we have the 
following relation with the signature ~$\chi$~:
\eqn\rrr{ \L(H_A) ~=~ \cases{ -\, n_A \ , & A=1,2,3 \cr 
-\, a_{A-4} \ , & A=5,...,N+4}} 
The minus signs are related to the fact that we work with 
lowest weight Verma modules (instead of the highest weight modules 
used in \Kab{}) and  to Verma module reducibility 
w.r.t. the roots ~$\a_A$ (this is explained in detail in \DPf). 

We have not fixed only the generator ~$H_4\,$. The standard
choice \Kab{} is a generator corresponding to the odd simple root
$\a_4\,$, but we can take any element of the Cartan subalgebra
which is not a linear combination of the established already
~$N+3$~ generators ~$H_A$. Our choice is to take the
generator ~$H_4$~ which corresponds 
to the root ~$\eps_3 + \eps_4$~ and which together with ~$\a_1,\a_2,\a_3$~ 
provides the root system of ~$so(8;\bbc)$.\foot{However, 
in the ~$osp(8/2N)$~ root system we have: ~$\eps_3 + \eps_4 
~=~ \a_3 + 2\a_4 + \cdots + 2\a_{N+3} + \a_{N+4}$.} 
The value ~$\L(H_4)$~ can not be a non-positive integer like 
the other ~$\L(H_A)$~ given in \rrr{}, 
since then we would obtain finite-dimensional representations 
of ~$so(8,\bbc)$~ \Kc, and thus, non-unitary representations of
~$so(6,2)$. In fact, unitarity w.r.t. ~$so(6,2)$~ would already
require that ~$\L(H_4)$~ is a non-negative number related
to the physically relevant conformal weight ~$d$, which is
related to the eigenvalue of the conformal Hamiltonian. 
That is why the lowest weight UIRs are also called ~{\it positive
energy} UIRs. We omit here the analysis by which it turns out that 
~$\L(H_4)$~ differs from ~$d$~ 
by the quantity ~$(n_1 + 2n_2 +n_3)/2$ (which is the value of the conformal 
Hamiltonian of the algebra ~$so(5,1)$ mentioned above). Thus, we set: 
\eqn\cftw{ \L(H_4) ~=~ d + \half (n_1 + 2n_2 +n_3) 
~=~ (\L , (\eps_3 + \eps_4)^\vee ) ~=~ (\L , \eps_3 + \eps_4)\ .}
This choice is consistent with the one in \Min, and the usage in 
\FSa. 

Having in hand the values of ~$\L$~ on the basis we can 
recover them for any element of ~$\ch$~ and ~$\ch^*$. 
In particular, for the values on the elementary functionals
we have from \rrr{} and \cftw{}: 
\eqna\vell
$$\eqalignno{ 
 (\L ,\eps_1 ) ~=&~ \half d\ 
-\ \quater (3 n_1 + 2 n_2 + n_3 ) &\vell{}\cr 
(\L ,\eps_2) ~=&~ \half d\ +\ \quater ( n_1 - 2 n_2 - n_3 )\cr 
(\L ,\eps_3) ~=&~ \half d\ +\ \quater ( n_1 + 2 n_2 - n_3 )\cr 
(\L ,\eps_4) ~=&~ \half d\ +\ \quater ( n_1 + 2 n_2 + 3 n_3 )\cr 
(\L ,\d_\hj) ~=&~ a_\hj + a_{\hj+1} + \cdots + \a_N ~\equiv~ r_\hj}$$ 
Using \rrr{} and \cftw{} one can write easily ~$\L = \L(\chi)$~ as a 
linear combination of the simple roots or of the 
elementary functionals ~$\eps_j\,,\d_\hj\,$, but this is 
not necessary in what follows. 

\subsec{Reducibility of Verma modules}

\nt 
Having established the relation between ~$\chi$~ and ~$\L$~ 
we turn our attention to the question of unitarity. 
The conditions of unitarity are intimately related with the 
conditions for  reducibility of the Verma modules w.r.t. to the 
odd positive roots. A Verma module ~$V^\L$~ is reducible w.r.t. 
the odd positive root ~$\g$~ iff the following holds \Kab:
\eqn\odr{ (\L - \r, \g) = 0 \ , \qquad \g \in \Delta^+_\I} 
where ~$\r\in\ch^*$~ is the very important in representation 
theory element given by the 
difference of the half-sums $\r_\o\,,\r_\I$
of the even, odd, resp., positive roots (cf. \psdm):
\eqn\hlf{ \eqalign{&\r ~\doteq~ \r_\o - \r_\I \cr 
&\r_\o\ =\ 3 \eps_1 + 2 \eps_2 + \eps_{3} + 
N \d_1 + (N-1) \d_2 + ... + 2\d_{N-1} + \d_N \cr 
&\r_\I \ =\ N (\eps_1 + \eps_2 + \eps_3 + \eps_4)}}

To make \odr{} explicit we need the values of ~$\L$~ 
and ~$\r$~ on the positive odd roots ~$\eps_i \pm \d_j$~ 
(which we obtain from \vell{}): 
\eqna\vodd
$$\eqalignno{ 
 (\L ,\eps_1 \pm\d_\hj ) ~=&~ \half d\ 
-\ \quater (3 n_1 + 2 n_2 + n_3 )\ \pm\ r_\hj 
&\vodd{a}\cr 
(\L ,\eps_2 \pm\d_\hj) ~=&~ \half d\ +\ \quater ( n_1 - 2 n_2 - n_3 )\ 
\pm\ r_\hj 
&\vodd{b}\cr 
(\L ,\eps_3\pm\d_\hj) ~=&~ \half d\ +\ \quater ( n_1 + 2 n_2 - n_3 )\ 
\pm\ r_\hj 
&\vodd{c}\cr 
(\L ,\eps_4\pm\d_\hj) ~=&~ \half d\ +\ \quater ( n_1 + 2 n_2 + 3 n_3 )\ 
\pm\ r_\hj 
&\vodd{d}\cr}$$
\eqn\ror{ (\,\r\, , \, \eps_i\, \pm\, \d_\hj\, ) ~=~ 
4-i-N \mp (N-\hj+1)}

Consecutively we find that the Verma module ~$V^{\L(\chi)}$~ 
is reducible if the conformal weight takes one of the 
following ~$8N$~ values ~$d^\pm_{ij}$~ 
labelled by the respective odd root ~$\eps_i \pm \d_\hj$~: 
\eqna\redd
$$\eqalignno{ 
 d ~=&~ d^\pm_{1\hj} ~\doteq~ \half (3 n_1 + 2 n_2 + n_3 ) + 2(3-N) 
\mp 2(r_\hj +N-\hj+1) &\redd {a}\cr
 d ~=&~ d^\pm_{2\hj} ~\doteq~ \half ( n_3 + 2 n_2 - n_1 ) + 2(2-N) 
\mp 2(r_\hj+N-\hj+1) &\redd {b}\cr
 d ~=&~ d^\pm_{3\hj} ~\doteq~ \half ( n_3 - 2 n_2 - n_1 ) + 2(1-N) 
\mp 2(r_\hj +N-\hj+1) &\redd {c}\cr
 d ~=&~ d^\pm_{4\hj} ~\doteq~ - \half ( n_1 + 2 n_2 + 3 n_3 ) -2N 
\mp 2(r_\hj +N-\hj+1) &\redd {d}\cr}$$
For future use we note the following relations:
\eqna\ordr
$$\eqalignno{
\half \( d^-_{i\hj} - d^-_{k\hl}\) ~=&~ n_i + \cdots + n_{k-1} +
k-i +\hl-\hj + a_\hj + \cdots + a_{\hl-1} ~>~ 0 \ , \cr & i\leq
k \,,\ \hj\leq\hl  \,,\ i\hj\neq k\hl &\ordr {a}\cr 
\half \( d^+_{i\hj} - d^+_{k\hl}\) ~=&~ n_i + \cdots + n_{k-1} +
k-i +\hj-\hl + a_\hl + \cdots + a_{\hj-1} ~>~ 0 \ , \cr & i\leq
k \,,\ \hj\geq\hl  \,,\ i\hj\neq k\hl &\ordr {b}\cr 
\half \( d^-_{i\hj} - d^+_{k\hl}\) ~=&~ 
n_i + \cdots + n_{k-1} + k-i + 2N-\hj-\hl 
+ r_\hj + r_\hl \ +\ 2 ~>~ 0 \ , \cr &
i\leq k &\ordr {c}}$$
which introduce some partial ordering between the 
quantities ~$d^\pm_{i\hj}$~ of which the essential 
would turn out to be the following:
\eqn\prrd{ d^-_{11} ~>~ d^-_{21} ~ >~ d^-_{31} ~>~ d^-_{41}}
The four values in \prrd{} play special role in the 
unitarity formulation. 
The value ~$d^-_{11}$~ is the biggest among all ~$d^\pm_{ij}$~
it is called 'the first reduction point' in \FF. 

\subsec{Shapovalov form and unitarity}

\nt
The Shapovalov form is a bilinear $\bbc$--valued form on Verma modules \Sha.
We need also the involutive antiautomorphism $\om$ of ~$U(\cg^+)$~ 
which will provide the real form we are interested in. 
Thus, an adaptation of the Shapovalov form suitable for our purposes 
is defined (as in \DPp{}) as follows:
\eqn\shh{ \eqalign{
&\(\ u\ ,\ u'\ \) ~~=~~
\(\ p\ v_0 \ ,\ p'\ v_0\ \) ~~\equiv~~
\(\ v_0 \ ,\ \om(p)\ p'\ v_0\ \)
~=~ \(\ \om(p')\ p\ v_0 \ , \ v_0\ \) ~,
\cr
&u ~=~ p\ v_0\ , ~ ~u' ~=~ p'\ v_0\ , \qquad p,p'\in U(\cg^+),
~~u,u' \in V^\L \cr}}
supplemented by the normalization condition ~$(v_0, v_0) ~=~ 1$.
 The norms squared of the states would be denoted by:
\eqn\nrm{ \| u \|^2 ~\equiv~ \(\ u\ ,\ u\ \) \ .}

We suppose that we consider representations which are unitary when 
restricted to the even part $\cg^+_\o$. This is justified
aposteriori since (as in the $D=4$ case \DPu,\DPp{}) the unitary
bounds of the even part are weaker than the supersymmetric ones
\FeFr. Thus, as in \DPu,\DPp{} 
we shall factorize the even part and we shall consider only the 
states created by the action of the odd generators, i.e., 
$\cf^\L = \( U(\cg^+)/U(\cg^+_\o)\) v_0$. We introduce notation ~$X^+_{i\hj}$~ 
for the odd generator corresponding to the positive root 
~$\eps_i - \d_\hj\,$, and ~$Y^+_{i\hj}$~ shall correspond to 
~$\eps_i + \d_\hj\,$. Since the odd generators 
are Grassmann there are only ~$2^{8N}$~ states in $\cf$ 
and choosing an ordering we give these states explicitly as follows:
\eqn\bas{ \eqalign{ 
\Psi_{{\bar \ve}{\bar \nu}} ~=&~ 
\( \ \prod_{i=1}^4\ (Y^+_{i1})^{\ve_{i1}} \)\ ...\ 
\( \ \prod_{i=1}^4\ (Y^+_{iN})^{\ve_{iN}} \)\ \times \cr 
&\times\ \( \ \prod_{i=1}^4\ (X^+_{iN})^{\nu_{iN}} \) 
\ ...\ \( \ \prod_{i=1}^4\ (X^+_{i1})^{\nu_{i1}} \) \, v_0 \ , \cr 
& \ve_{i\hj},\nu_{i\hj} = 0,1}}
where ~${\bar \ve}$, ${\bar \nu}$, 
~ denote the set of all ~$\ve_{i\hj}\,$, ~$\nu_{i\hj}\,$, resp.
For future use we give 
notation for the number of $Y$'s and $X$'s:
\eqn\nnn{ \ve ~\equiv~ \sum_{i=1}^4 \sum_{\hj=1}^N \ve_{i\hj}\ , \quad 
\nu ~\equiv~ \sum_{i=1}^4 \sum_{\hj=1}^N \nu_{i\hj}\ ,} 
and through them for the ~{\it level}~ $\ell$~:
\eqn\levv{ \ell \(\Psi_{{\bar \ve}{\bar \nu}}\) ~=~ \ve+\nu ~.}

\subsec{Explicit realization of the basis of\ osp(8/2N)} 

\nt 
To proceed further we need the explicit realization of 
the generators of ~$osp(8/2N)$. It is obtained from the 
standard one of \Kab{} by applying a unitary transformation 
done in order to bring the Cartan subalgebra in diagonal form. 
The matrices are ~$(8+2N)\times(8+2N)$ and are 
in standard supermatrix form, i.e., the even ones are 
of the form:
$$\pmatrix{* & 0\cr 0 & *\cr}$$
and the odd ones of the form: 
$$\pmatrix{0 & *\cr * & 0\cr}$$
The description is done very conveniently in terms of the matrices 
~$E_{AB}\in gl(8/2N,\bbc)$, $A,B =1,...,8+2N$. 
Fix ~$A,B$, then the matrix 
~$E_{AB}$~ has only non--zero entry, equal to 1, at the
intersection of the $A$-th row and $B$-th column. 

Then the generators ~$H_A$~ are given by: 
\eqn\bash{ 
\eqalign{
 H_j ~=&~ E_{jj} - E_{j+1,j+1} -
 E_{j+4,j+4} + E_{j+5,j+5} \ , \quad j=1,2,3 \cr 
H_4 ~=&~ E_{33} + E_{44} - E_{77}- E_{88} \cr 
H_{4+\hj} ~=&~ E_{8+\hj,8+\hj} - E_{9+\hj,9+\hj} -  E_{8+N+\hj,8+N+\hj} +
E_{9+N+\hj,9+N+\hj}\ , \quad \hj=1,...,N-1 \cr
H_{4+N} ~=&~ E_{8+N,8+N} - E_{8+2N,8+2N}}}
 the basis of $\cg^+$ - enumerated by the corresponding roots - is:
\eqna\psr 
$$\eqalignno{ 
& L^+_{ij} \ = \ E_{ij} - E_{4+j,4+i} \ , 
\quad {\rm roots:}\ \eps_i - \eps_j ~, \quad i<j\cr 
& P^+_{ij} \ = \ E_{i,4+j} - E_{j,4+i} 
\ , \quad {\rm roots:}\ \eps_i + \eps_j ~, \quad i<j\cr
& T^+_{\hi\hj} \ = \ E_{8+\hi,8+\hj} - E_{8+N+\hj,8+N+\hi} 
\ , \quad {\rm roots:}\ \d_\hi - \d_\hj ~, \quad \hi<\hj\cr 
& R^+_{\hi\hj} \ = \ E_{8+\hi,8+N+\hj} + E_{8+\hj,8+N+\hi} 
\ , \quad {\rm roots:}\ \d_\hi + \d_\hj ~, \quad \hi<\hj\cr 
& R^+_{\hi} \ = \ E_{8+\hi,8+N+\hi} 
\ , \quad {\rm roots:}\ 2\d_\hi \cr 
& X_{i\hj}^+ \ = \ E_{i,8+\hj} +  E_{8+N+\hj,4+i}  
\ , \quad {\rm roots:}\ \eps_i-\d_\hj  \cr
& Y_{i\hj}^+ \ = \ E_{i,8+N+\hj} -  E_{8+\hj,4+i}  
\ , \quad {\rm roots:}\ \eps_i+\d_\hj &\psr{}}$$
while the basis of $\cg^-$ is:
\eqna\ngr 
$$\eqalignno{ 
& L^-_{ij} \ = \ E_{ij} - E_{4+j,4+i} \ , 
\quad {\rm roots:}\ \eps_i - \eps_j ~, \quad i>j\cr 
& P^-_{ij} \ = \ E_{4+j,i} - E_{4+i,j}
\ , \quad {\rm roots:}\ -(\eps_i + \eps_j) ~, \quad i<j\cr
& T^-_{\hi\hj} \ = \ E_{8+\hi,8+\hj} - E_{8+N+\hj,8+N+\hi} 
\ , \quad {\rm roots:}\ \d_\hi - \d_\hj ~, \quad \hi>\hj\cr 
& R^-_{\hi\hj} \ = \ E_{8+N+\hi,8+\hj} + E_{8+N+\hj,8+\hi} 
\ , \quad {\rm roots:}\ -(\d_\hi + \d_\hj) ~, \quad \hi<\hj\cr 
& R^-_{\hi} \ = \ E_{8+N+\hi,8+\hi} 
\ , \quad {\rm roots:}\ -2\d_\hi \cr 
& X_{i\hj}^- \ = \ E_{4+\hi,8+N+\hj} -  E_{8+\hj,i}  
\ , \quad {\rm roots:}\ -\eps_i+\d_\hj  \cr
& Y_{i\hj}^- \ = \ E_{4+i,8+\hj} +  E_{8+N+\hj,i}  
\ , \quad {\rm roots:}\ -(\eps_i+\d_\hj) &\ngr{}}$$

{}From the explicit matrix realization above one easily obtains 
all commutation relations. We shall write down only some more 
important ones:
\eqna\odcr
$$\eqalignno{ 
&[\, X_{i\hj}^+ \, ,\, X_{i\hj}^-\,]_+ ~=~ - E_{ii} + E_{4+i,4+i} 
- E_{8+\hj,8+\hj} + E_{8+N+\hj,8+N+\hj} ~=~ - \hH_i -
\tH_\hj\quad &\odcr {a}\cr 
&[\, Y_{i\hj}^+ \, ,\, Y_{i\hj}^-\,]_+ ~=~ E_{ii} - E_{4+i,4+i} 
- E_{8+\hj,8+\hj} + E_{8+N+\hj,8+N+\hj} ~=~ \hH_i - \tH_\hj
&\odcr {b}\cr}$$ 
where we have introduced notation for an alternative basis of ~$\ch$~ 
which actually is used in the calculation of scalar products:
$$\eqalignno{ &\hH_i ~\equiv~ E_{ii} - E_{4+i,4+i} \ , 
\quad i=1,...,4 &\odcr {c}\cr 
&\tH_\hj ~\equiv~ E_{8+\hj,8+\hj} - E_{8+N+\hj,8+N+\hj}\ , 
\quad \hj=1,...,N &\odcr {d}\cr}$$
In particular, we shall use continuously:
\eqna\edcr
$$\eqalignno{ 
&[\, \hH_k \, ,\, X_{i\hj}^+\,] ~=~ \d_{ki}\, X_{i\hj}^+ &\edcr {a}\cr
&[\, \hH_k \, ,\, Y_{i\hj}^+\,] ~=~ \d_{ki}\, Y_{i\hj}^+ &\edcr {b}\cr
&[\, \tH_\hl \, ,\, X_{i\hj}^+\,] ~=~ -\d_{\hl \hj}\, X_{i\hj}^+ &\edcr {c}\cr
&[\, \tH_\hl \, ,\, Y_{i\hj}^+\,] ~=~ \d_{\hl \hj}\, Y_{i\hj}^+
&\edcr {d}\cr}$$ 

We also give the generators $\hH_A$ in terms of $H_A$ 
\eqn\nebb{ \eqalign{
\hH_1 ~=&~ H_1 + H_2 + \half (H_4 + H_3)  \cr 
\hH_2 ~=&~ H_2 + \half (H_4 + H_3)  \cr 
\hH_3 ~=&~ \half (H_4 + H_3)  \cr 
\hH_4 ~=&~ \half (H_4 - H_3)  \cr 
\tH_\hj ~=&~ H_{4+\hj} + \cdots + H_{4+N} \ , \quad \hj=1,...,N \ .}}

\newsec{Unitarity} 

\nt 
In this Section we state our main result (in the Theorem) 
on the lowest weight (positive energy) UIRs and give 
the Proof of necessity in general and Proof of sufficiency 
at generic points (the reduction points are dealt with 
in the next Section). 

\subsec{Calculation of some norms}

\nt 
In this subsection we show how to use the form \shh{} to 
calculate the norms of the states from ~$\cf$. 

We first need explicitly the conjugation 
~$\om$~ on the odd generators: 
\eqn\con{ \om \(X^+_{i\hj}\) ~ = ~ - X^-_{i\hj} \ , \qquad  
\om \(Y^+_{i\hj}\) ~ = ~ Y^-_{i\hj}}
(In matrix notation this would follow from: $\om (E_{i,8+\hj}) = 
E_{8+\hj,i}\,$, $\om (E_{i+4,8+\hj}) = - E_{8+\hj,i+4}\,$.)

We give now explicitly the norms of the one-particle states from 
~$\cf$~ introducing also notation for future use:
\eqna\norm
$$\eqalignno{
x_{i\hj} ~~\equiv& ~~ \|\, X^+_{i\hj}\  v_0\, \|^2 ~=~ 
\(\, X^+_{i\hj} \ v_0 \ ,\ X^+_{i\hj}\ v_0\, \) ~=\cr 
=&\ - \(\ v_0 \ ,\ X^-_{i\hj}\,X^+_{i\hj}\ v_0\ \) \ = \ 
\(\ v_0 \ ,\ (\hH_i+\tH_\hj) \ v_0\ \) ~=\cr
=&\ \L (\hH_i+\tH_\hj) &\norm{a} \cr 
y_{i\hj} ~~\equiv& ~~ \|\, Y^+_{i\hj} \ v_0\, \|^2 ~=~ 
\(\ Y^+_{i\hj} \ v_0 \ ,\ Y^+_{i\hj}\ v_0\ \) ~=\cr 
=&\ \(\ v_0 \ ,\ Y^-_{i\hj}\,Y^+_{i\hj}\ v_0\ \) \ = \ 
\(\ v_0 \ ,\ (\hH_i -\tH_\hj) \ v_0\ \) ~=\cr
=&\ \L (\hH_i-\tH_\hj) &\norm{b}}$$
Using \nebb{}, \rrr{} and \cftw{}  we get: 
\eqna\normm
$$\eqalignno{
x_{i\hj} ~=&~ \( \L\ ,\ \eps_i - \d_\hj \) ~=~ \half
(d-d^-_{i\hj}) + 5-i-\hj 
&\normm{a}\cr 
y_{i\hj} ~=&~ \( \L\ ,\ \eps_i + \d_\hj \) ~=~ \half
(d-d^+_{i\hj}) + 3-i+\hj-2N 
&\normm{b}}$$
And we note:
\eqna\comp
$$\eqalignno{ 
&x_{i+1,\hj} - x_{i\hj} ~=~ \half ( d^-_{i\hj} - d^-_{i+1,\hj}) -
1 ~=~ n_i ~\geq 0 
\ , &\comp{a}\cr 
&x_{i,\hj+1} - x_{i\hj} ~=~ \half ( d^-_{i\hj} - d^-_{i,\hj+1}) -
1 ~=~ a_\hj ~\geq 0 
\ , &\comp{b}\cr 
&y_{i+1,\hj} - y_{i\hj} ~=~ \half ( d^+_{i\hj} - d^+_{i+1,\hj}) -
1 ~=~ n_i ~\geq 0 \ , &\comp{c}\cr 
&y_{i\hj} - y_{i,\hj+1} ~=~ \half ( d^+_{i,\hj+1} - d^+_{i\hj}) -
1 ~=~ a_\hj 
~\geq 0 \ , &\comp{d}\cr 
&y_{i,\hl} - x_{i\hj} ~=~ \half ( d^-_{i\hj} - d^+_{i,\hl}) +
\hj+\hl-2N -2 ~=~ r_\hl + r_\hj ~\geq 0 \ , &\comp{e}\cr}$$
Thus, ~$x_{11}$~ is the smallest among all ~$x_{i\hj}$~ and ~$y_{i\hj}\,$. 

\subsec{Statement of main result and proof of necessity}

\nt 
In this subsection we state our main result in the Theorem and give 
the Proof of necessity via two Propositions (1 \& 2). 

First we give the norms which actually determine all of the unitarity 
conditions. In order to simplify the exposition we shall use also the 
notation:
\eqn\newn{ \eqalign{ &X^+_j ~\equiv~ X^+_{j1} \cr 
&x_j ~\equiv~ x_{j1} \cr}}
We note in these terms a subset of \comp{a}
$$\eqalignno{ 
&x_{i+1} - x_{i} ~=~ n_i ~\geq 0 
\ , &\comp{a'}\cr}$$

Next we calculate: 
\eqna\nrmx
$$\eqalignno{
& \|\, X^+_j\ X^+_k\ v_0\, \|^2 ~=~ (x_j-1) x_k \ , \quad j<k &\nrmx a\cr 
& \|\, X^+_j\ X^+_k\ X^+_\ell\ v_0\, \|^2 ~=~ (x_j-2) (x_k-1) x_\ell \ , 
\quad j<k< \ell &\nrmx b\cr
& \|\, X^+_1\ X^+_2\ X^+_3\ X^+_4 \ v_0\, \|^2 ~=~ 
(x_1-3) (x_2-2) (x_3-1) x_4 \ , &\nrmx c\cr}$$ 
 The norms \norm{a} and \nrmx{} 
are all strictly positive iff ~$x_j > 4-j$, $j=1,2,3,4$, 
which are all fulfilled if~ $x_1 ~>~ 3 $, since ~$x_1$~ is the smallest 
among the ~$x_j$. Thus, the these norms are strictly 
positive iff:
\eqn\suff{ x_1 ~>~ 3 ~~ \Longleftrightarrow ~~ d ~>~ d^-_{11} \ .} 
It turns out that this restriction is sufficient 
to guarantee unitarity of the whole representation. 
This is not unexpected: in all cases studied so far 
it was always so that if ~$d$~ is bigger than the 
first odd reduction point then the module is unitary. 

Of course, the condition \suff{} is not necessary for unitarity. 
On the experience so far it is expected that when ~$d$~ is equal 
to some of the reducibility values then unitarity is also 
possible, though in these cases there would be some conditions 
on the representation parameters, and one 
has to factor out the resulting zero norm states. Now 
we can formulate the main result: 

\nt {\bf Theorem:}~~ All positive energy unitary irreducible 
representations of the 
conformal superalgebra ~$osp(8^*/2N)$~ characterized by the signature 
~$\chi$~ in \sgn{} are obtained for real ~$d$~ and are given 
in the following list: 
\eqna\unt 
$$\eqalignno{
&d ~\geq~ d^-_{11} ~=~ \half (3 n_1 + 2 n_2 + n_3 ) + 2r_1 +6 
\ , \quad {\rm no\ restrictions\ on}\ n_j &\unt a\cr 
&d ~=~ d^-_{21} ~=~ \half ( n_3 + 2 n_2 )+ 2r_1 + 4 
\ , \quad n_1 =0 &\unt b\cr
&d ~=~ d^-_{31} ~=~ \half n_3 + 2r_1 + 2 
\ , \quad n_1 =n_2 = 0 &\unt c\cr
&d ~=~ d^-_{41} ~=~ 2r_1 
\ , \quad n_1 =n_2 = n_3 = 0 &\unt d\cr}$$

\nt {\it Remark:}~~
For ~$N=1,2$~ the Theorem was conjectured by Minwalla \Min{}, 
except that he conjectured unitarity also for the open 
interval $(d^-_{31},d^-_{21})$ with conditions on $n_j$ as in \unt{c}. 
We should note that this conjecture could be reproduced neither by
methods of conformal field theory \FSa, nor by the 
oscillator method \GSB{} (cf. \Min), and thus was in doubt.  
To compare with the notations of \Min{} one should use the 
following substitutions: $n_1\ =\ h_2-h_3\,$, $n_2\ =\ h_1-h_2\,$, 
 $n_3\ =\ h_2+h_3\,$, $r_1=k$, and $h_j$ are all integer or 
all half-integer. The fact that $n_j\geq 0$ 
for $j=1,2,3$ translates into: ~$h_1 \geq h_2 \geq |h_3|$, i.e., 
the parameters $h_j$ are of the type often used for 
representations of $so(2N)$ (though usually for $N\geq 4$).
Note also that the statement of the Theorem is arranged in
\Min{} according to the possible values of ~$n_i$~ first and 
then the possible values of ~$d$. To compare with the notation of
\FSa{} we use the substitution ~$(n_1,n_2,n_3)\to (J_3,J_2,J_1)$.
Some UIRs at the four exceptional points ~$d^-_{i1}$~ were constructed
in \GNWT{} by the oscillator method (some of these were identified
with Cartan-type signatures like \sgn{} in, e.g., \Min, \EFS).\dia 

The ~{\bf Proof}~ of the Theorem requires to show that there is unitarity 
as claimed, i.e., that the conditions are ~{\it sufficient}~ 
and that there is no unitarity otherwise, i.e., that the conditions 
are ~{\it necessary}. For the sufficiency  
we need all norms, but for the necessity part we need 
only the knowledge of a few norms. We give the necessity 
part in two Propositions.

\pr{1} There is no unitarity in any of the 
open intervals: $(d^-_{j+1,1},d^-_{j1})$, $j=1,2,3$, 
and if ~$d<d^-_{41}\,$. \nl 
{\it Proof:} 
\item{\bu} Consider ~$d$~ in the open interval ~$(d^-_{21},d^-_{11})$, which 
means that ~$3 > x_1 > 2- n_1\,$. Consider the norm \nrmx{c}
and using \comp{a} express all $x_i$ in terms of $x_1\,$. We have:
\eqn\prfa{ (x_1-3) (x_2-2) (x_3-1) x_4 ~=~ (x_1-3) (x_1 + n_1 -2) 
(x_1 +n_1 +n_2 -1)(x_1 +n_1 +n_2 + n_3 )}
The first term is strictly negative while the other three terms 
are strictly positive, independently of the values of $n_i\,$.
Thus, the norm \nrmx{c} is negative in the open interval 
~$(d^-_{21},d^-_{11})$. 
\item{\bu} Consider ~$d$~ in the open interval ~$(d^-_{31},d^-_{21})$, which 
means that ~$2 > x_1 + n_1 > 1- n_2\,$. Consider the norm \nrmx{b} 
for $(j,k,\ell) = (1,3,4)$ 
and using \comp{a} express all $x_i$ in terms of $x_1\,$. We have:
\eqn\prfb{ (x_1-2) (x_3-1) x_4 ~=~ (x_1-2) 
(x_1 +n_1 +n_2 -1)(x_1 +n_1 +n_2 + n_3 )}
The first term is strictly negative while the other two terms 
are strictly positive, independently of the values of $n_i\,$.
Thus, the norm of the state ~$X^+_1\, X^+_3\, X^+_4\, v_0$~ 
is negative in the open interval ~$(d^-_{32},d^-_{21})$. 
\item{\bu} Consider ~$d$~ in the open interval ~$(d^-_{41},d^-_{31})$, which 
means that ~$1 > x_1 + n_1 +n_2 > - n_3\,$. Consider the norm \nrmx{a} 
for $(j,k) = (1,4)$ 
and using \comp{a} express all $x_i$ in terms of $x_1\,$. We have:
\eqn\prfc{ (x_1-1) x_4 ~=~ (x_1-1) (x_1 +n_1 +n_2 + n_3 )}
The first term is strictly negative while the second
is strictly positive, independently of the values of $n_i\,$.
Thus, the norm of the state ~$X^+_1\, X^+_4\, v_0$~ 
is negative in the open interval ~$(d^-_{31},d^-_{21})$. 
\item{\bu} Consider ~$d$~ in the infinite open interval ~$d<d^-_{41}\,$. 
Then the norm of ~$X^+_{41}\,v_0$~ is negative using \normm{a}:
$$ x_4 ~=~ x_{41} ~=~ \half (d-d^-_{41}) ~<~ 0 \ .$$ 
Thus, the Proposition is proved.\bsq

Thus, we have shown the exclusion of the open intervals in 
the statement of the Theorem. It remains to show the 
necessity of the restrictions on ~$n_i$~ in cases ~$b,c,d$~ 
of the Theorem.

\pr{2} There is no unitarity in the following cases:
$$\eqalignno{
&d ~=~ d^-_{21} 
\ , \quad n_1 > 0 &\unt {b'}\cr
&d ~=~ d^-_{31} 
\ , \quad n_1 + n_2 > 0 &\unt {c'}\cr
&d ~=~ d^-_{41} 
\ , \quad n_1 + n_2 +n_3 > 0 &\unt {d'}\cr}$$
{\it Proof:} 
\item{\bu}
Let ~$d ~=~ d^-_{21}$~ which means ~$x_2=2$~ and ~$x_1=2-n_1\,$. 
Consider again the norm of ~$X^+_1\, X^+_3\, X^+_4\, v_0$~ and substitute 
the value of ~$x_1$~ in \prfb{} to get:
\eqn\prfbb{ (x_1-2) (x_3-1) x_4 ~=~ (-n_1) (1+n_2 )(2+n_2 + n_3 )}
This norm is negative if ~$n_1>0$. 
\item{\bu}
Let ~$d ~=~ d^-_{31}$~ which means ~$x_3=1$~ and ~$x_1=1-n_1-n_2\,$.
Consider again the norm of ~$X^+_1\, X^+_4\, v_0$~ and substitute 
the value of ~$x_1$~ in \prfc{} to get:
\eqn\prfbb{ (x_1-1) x_4 ~=~ (-n_1-n_2) (1+n_3 )} 
This norm is negative if ~$n_1+n_2>0$.
\item{\bu}
Let ~$d ~=~ d^-_{41}$~ which means ~$x_4=0$~ and ~$x_1=-n_1-n_2-n_3\,$.
But the latter is the norm of ~$X^+_1\ v_0$~ and it is 
negative if ~$n_1 + n_2 +n_3 > 0$.\nl 
Thus, the Proposition is proved.\bsq 

With this we have shown that the conditions of the Theorem are 
~{\it necessary}.\bsq 

The proof of sufficiency is postponed for the next subsection. 

\re{2} The reader may wonder why the other reducibility points
are not playing such an important role as the quartet appearing
in the Theorem.

First we note that the analogous calculations involving other quartets 
of operators: ~$X^+_{i\hj}$~ ($\hj\neq 1$ fixed, $i=1,2,3,4$) 
or ~$Y^+_{i\hj}$~ ($\hj$ fixed, $i=1,2,3,4$) 
give the same results as 
\nrmx{} with just replacing ~$x_i \to x_{i\hj}$~ or ~$x_i \to y_{i\hj}$. 
This brings the conditions ~$x_{i\hj} > 4-i$~ or ~$y_{i\hj} > 4-i$~ 
which all follow from ~$x_1 > 3$~ because of \comp{}.  
This is related to the fact that ~$d^-_{11}$~ is the largest 
reduction point. 

Further, we may look for the analog of Proposition 1 and we can 
prove the same results involving ~$d^-_{i\hj}\,$, ($\hj\neq 1$ fixed),
or  ~$d^+_{i\hj}\,$, ($\hj$ fixed). However, these results 
may be relevant only if the exceptional points ~$d^-_{21}\,$, 
~$d^-_{31}\,$, ~$d^-_{41}\,$, together with the respective 
conditions, would happen to be in some of the open intervals 
defined by some other quartet, which would prove their non-unitarity. 
The reason that this does not happen is the following. 

Let ~$n_1 ~=~ 0$. Then one can easily see that ~$d^-_{11}$~ and
~$d^-_{21}$~ are the two largest reduction points (for $N=1,2$
cf. \Min{}), i.e., 
all other reduction points are smaller, and thus, ~$d^-_{21}$~
can not be in any open interval defined by some other quartet.

Analogously, for $n_1 ~=~ n_2 ~=~ 0$~ one can easily see that
~$d^-_{11}\,$, ~$d^-_{21}$~ and ~$d^-_{31}$~ are the three
largest reduction points (for $N=1,2$
cf. \Min{}), and so ~$d^-_{31}$~ can not be in
any open interval defined by some other quartet. Finally, for
$n_1 ~=~ n_2 ~=~ n_3 ~=~ 0$~ the points ~$d^-_{11}\,$,
~$d^-_{21}\,$, ~$d^-_{31}$~ and ~$d^-_{41}$~ are the four
largest reduction points (for $N=1,2$
cf. \Min{}), and ~$d^-_{41}$~ can not be in
any open interval defined by some other quartet.\dia

\subsec{General form of the norms and unitarity in the generic
case} 

\nt 
In this subsection we give the Proof of sufficiency of the 
Theorem in the generic case. This requires the general form 
of the norms. The states are divided into classes and the 
norms are given for the different cases in Propositions (3-7).
At the end we finish the Proof of sufficiency utilizing these 
Propositions.

To present the general formulae for the norms 
we first we divide the states into factorizable  
and unfactorizable as follows. Let the first generator in
~$\Psi_{{\bar \ve}{\bar\nu}}$~ be ~$Y^+_{i\hj}\,$, i.e.,
~$\Psi_{{\bar \ve}{\bar 
\nu}} ~=~ Y^+_{i\hj}\, ...\, v_0\,$. Then ~$\Psi_{{\bar \ve}{\bar
\nu}}$~ is called ~{\it factorizable}~ if the following
three statements hold:
\eqna\facty
$$\eqalignno{ 
&\eps_{k\hj}\, \eps_{i\hl} ~=~ 0 ~~{\rm or}~~ \eps_{k\hl} ~=~ 1 \ 
, \quad {\rm for\ all\ pairs} \
(k,\hl)\ {\rm such\ that:}\ k>i,\ \hl >\hj &\facty {a}\cr 
&\eps_{k\hj}\, \nu_{i\hl} ~=~ 0 ~~{\rm or}~~ \nu_{k\hl} ~=~ 1 \ 
, \quad {\rm for\ all} \ k>i,\
{\rm and\ all}\ \hl &\facty {b}\cr 
&\eps_{i\hl}\, \nu_{j \hl} ~=~ 0 ~~{\rm or}~~ \nu_{j \hj} ~=~ 1 \ 
,\quad {\rm for\ all} \ \hl>\hj,\
{\rm and\ all}\ j &\facty {c}\cr}$$
If the first generator in ~$\Psi_{{\bar \ve}{\bar\nu}}$~ is ~$X^+_{i\hj}\,$,
so that ~${\bar\ve} =0$, 
then ~$\Psi_{0,{\bar\nu}}$~ is called ~{\it factorizable}~ 
if the following statement holds:
\eqn\factx{\nu_{j\hj}\, \nu_{i\hl} ~=~ 0 ~~{\rm or}~~ \nu_{j\hl} ~=~ 1 \ 
, \quad {\rm for\ all\ pairs} \
(j,\hl)\ {\rm such\ that:}\ j>i,\ \hl <\hj}

\nt 
Our first result on the norms is:\nl 
\pr{3}
For factorizable states starting with ~$X^+_{i\hj}$~ 
the following relation holds: 
\eqna\ffx
$$\eqalignno{ 
\|\, \Psi_{0,{\bar \nu}}\, \|^2 ~=&~ 
\(x_{i\hj} + {\tnu}_{i\hj} \)\ 
\|\, \Psi_{0,{\bar \nu'}}\, \|^2 &\ffx {}\cr 
& {\tnu}_{i\hj} ~=~ \nu_{i,\hj-1} + \cdots + \nu_{i,1} -
\nu_{i+1,\hj} - \cdots - \nu_{4,\hj} \cr 
&\nu'_{j\hl} =\nu_{j\hl} - \d_{ji}\d_{\hl \hj}}$$
For factorizable states starting with ~$Y^+_{i\hj}$~ 
the following relation holds: 
\eqna\ffy
$$\eqalignno{ 
\|\, \Psi_{{\bar\ve},{\bar \nu}}\, \|^2 ~=&~ 
\(y_{i\hj} + {\tilde \ve}_{i\hj} + \nu_i + {\hnu}_\hj \)\ 
\|\, \Psi_{{\bar \ve'}{\bar \nu}}\, \|^2 &\ffy {}\cr 
& {\tilde \ve}_{i\hj} ~=~ \ve_{i,\hj+1} + \cdots + \ve_{i,N} - 
\ve_{i+1,\hj} - \cdots - \ve_{4,\hj} \cr 
&\nu_i ~=~ \nu_{i,1} + \cdots + \nu_{i,N} \ , \quad 
{\hnu}_\hj ~=~ \nu_{1,\hj} + \cdots + \nu_{4,\hj} \cr 
&\ve'_{j\hl} =\ve_{j\hl} - \d_{ji}\d_{\hl \hj}}$$
{\it Proof:}~~ We start with \ffx. Clearly, 
~$\Psi_{0,{\bar \nu}} ~=~ X^+_{i\hj}\, \Psi_{0,{\bar \nu'}}$. 
Then the norm squared is:
\eqna\nrmxx
$$\eqalignno{\|\, \Psi_{0,{\bar \nu}}\, \|^2 ~=&~
\( X^+_{i\hj}\, \Psi_{0,{\bar \nu'}}\ ,\ X^+_{i\hj}\,
\Psi_{0,{\bar \nu'}} \) ~=~ - \(\Psi_{0,{\bar \nu'}}\ ,\
X^-_{i\hj}\, X^+_{i\hj}\, \Psi_{0,{\bar \nu'}}\) ~=&\nrmxx{a}\cr 
=&~ - \(\Psi_{0,{\bar \nu'}}\ ,\ \( - X^+_{i\hj}\, 
X^-_{i\hj}\, - \hH_i-\tH_\hj \) 
 \Psi_{0,{\bar \nu'}}\) ~=&\nrmxx{b}\cr 
=&~ \(\Psi_{0,{\bar \nu'}}\ ,\ \(  \hH_i+\tH_\hj \) 
 \Psi_{0,{\bar \nu'}}\) ~=&\nrmxx{c}\cr 
= &~\( \L ( \hH_i+\tH_\hj) + {\tnu}_{i\hj} \) \(\Psi_{0,{\bar
\nu'}}\ ,\ \Psi_{0,{\bar \nu'}}\) ~=&\nrmxx{d}\cr 
=&~ \(x_{i\hj} + {\tnu}_{i\hj} \)\ 
\|\, \Psi_{0,{\bar \nu'}}\, \|^2 \ , &\nrmxx{e}\cr}$$
Note that the term ~$X^+_{i\hj}\, X^-_{i\hj}$~ in \nrmxx{b} gives
no contribution: due to conditions \factx{} the operator
~$X^-_{i\hj}$~ anticommutes with the operators in
~$\Psi_{0,{\bar \nu'}}$~ or produces terms like: ~$(X^+_{j\hl})^2
~=~ 0$, thus it reaches ~$v_0$~ without additional terms. 
Moving the operator ~$\hH_i+\tH_\hj$~ through ~$\Psi_{0,{\bar
\nu'}}$~ produces the addition ~${\tnu}_{i\hj}$~ - the terms 
~$\nu_{i,\hj-1} + \cdots + \nu_{i,1}$~ are due to \edcr{a}, 
and the terms ~$-\nu_{i+1,\hj} - \cdots - \nu_{4,\hj}$~ 
are due to \edcr{c}.
Analogously, we consider \ffy{}. Clearly, 
~$\Psi_{{\bar\ve},{\bar \nu}} ~=~ Y^+_{i\hj}\,
\Psi_{{\bar\ve'},{\bar \nu}}$. 
The norm squared is:
\eqna\nrmxx
$$\eqalignno{\|\, \Psi_{{\bar \ve}{\bar \nu}}\, \|^2 ~=&~
\(Y^+_{i\hj}\, \Psi_{{\bar\ve'},{\bar \nu}}\ ,\ Y^+_{i\hj}\,
\Psi_{{\bar\ve'},{\bar \nu}} \) ~=~ 
\( \Psi_{{\bar\ve'},{\bar \nu}}\ ,\ Y^-_{i\hj}\, Y^+_{i\hj}\, 
 \Psi_{{\bar\ve'},{\bar \nu}}\) ~=&\nrmxx{a}\cr 
=&~ \( \Psi_{{\bar\ve'},{\bar \nu}}\ ,\ \( - Y^+_{i\hj}\, 
Y^-_{i\hj} +  \hH_i-\tH_\hj \) 
 \Psi_{{\bar\ve'},{\bar \nu}}\) ~=&\nrmxx{b}\cr 
=&~ \( \Psi_{{\bar\ve'},{\bar \nu}}\ ,\ \(  \hH_i-\tH_\hj \) 
 \Psi_{{\bar\ve'},{\bar \nu}}\) ~=&\nrmxx{c}\cr 
= &~\( \L ( \hH_i-\tH_\hj) + {\tilde \ve}_{i\hj} + \nu_i +
{\hnu}_\hj \) \( \Psi_{{\bar\ve'},{\bar \nu}} \ ,\  
 \Psi_{{\bar\ve'},{\bar \nu}}\) ~=&\nrmxx{d}\cr 
=&~ \(y_{i\hj} + {\tilde \ve}_{i\hj} + \nu_i + {\hnu}_\hj \)\ 
\|\, \Psi_{{\bar\ve'},{\bar \nu}}\, \|^2 \ , &\nrmxx{e}\cr}$$
Note that to produce the additional terms ~$ {\tilde \ve}_{i\hj}
+ \nu_i + {\hnu}_\hj$~ we need all of \edcr{}.\bsq 

The states ~$\Psi_{{\bar \ve'}{\bar \nu}}$~ and
~$\Psi_{0,{\bar\nu'}}$~ may still be factorizable and so on. The
state ~$\Psi_{0,{\bar \nu}}$~ is called ~{\it fully
factorizable}~ if the process of factorization can be repeated
~$\nu$~ times. The state ~$\Psi_{{\bar\ve},{\bar \nu}}$~ is
called ~{\it fully factorizable}~ if the process 
of factorization can be repeated ~$\ve$~ times and the 
resulting state ~$\Psi_{0,{\bar \nu}}$~ is fully factorizable. 

\nt 
Our first main result on the norms is:\nl 
\pr{4} The norm of a fully factorizable state 
 ~$\Psi_{{\bar \ve}{\bar \nu}}$~ 
is given by the following formula:
\eqna\nrms
$$\eqalignno{
\|\, \Psi_{{\bar \ve}{\bar \nu}}\, \|^2 ~~=
&~~ \cn_{{\bar \ve}{\bar \nu}} \ , &\nrms {}}$$
where 
\eqna\nrmss
$$\eqalignno{
\cn_{{\bar \ve}{\bar \nu}} ~=&~ 
\prod_{i=1}^4 \ \prod_{\hj=1}^N \ 
\( y_{i\hj} +{\tilde \ve}_{i\hj} + \nu_i + {\tnu}_\hj
\)^{\ve_{i\hj}}\  
\( x_{i\hj} +{\tnu}_{i\hj} \)^{\nu_{i\hj}} 
&\nrmss {}}$$
{\it Proof:}~~ By direct iteration of \ffy{} and \ffx{}.\bsq 

Naturally, the norms in \nrmx{} are special cases of \nrms{}. 

Note that the norm squared of a state is a polynomial in ~$d$~ 
of degree the level ~$\ell$~ of the state. 

We shall now discuss states which are not fully factorizable. 
It is enough to consider unfactorizable states, since if a state
is factorizable then we apply \ffy{} or \ffx{} one or more times
until we are left with the norm squared of an unfactorizable
state. We shall have two propositions, the first of which is:

\pr{5} Let $\Psi_{0,{\bar \nu}}$ be an unfactorizable state
starting with the generator ~$X^+_{i\hj}\,$.
This means that there are one or more pairs
of integers ~$(k,\hl)$~ so that \factx{} is violated. 
Let us enumerate the pairs violating \factx{} as:
\eqn\enux{ (k_m,\hl_{m,n})\ , \quad i < k_1 < ... < k_p 
\ , \quad \hj > \hl_{m,1} > ... > \hl_{m,q(m)}} 
so that the following holds:
\eqn\fatx{ \nu_{k_m,\hj}~=~ \nu_{i,\hl_{m,n}} ~=~ 1 ~~{\rm and}~~
\nu_{k_m,\hl_{m,n}} ~=~ 0} 
Then the norm of ~$\Psi_{0,{\bar \nu}}$~ 
is given by the following formula:
\eqna\nrmg
$$\eqalignno{
\|\, \Psi_{0,{\bar \nu}}\, \|^2 ~~=&~~
 \(x_{i,\hj} +{\tnu}_{i,\hj} \)\ 
\|\, \Psi_{0,{\bar \nu'}}\, \|^2 ~-~ \sum_{m=1}^p\
\sum_{n=1}^{q(m)}\ \cc^{m,n}_{0,{\bar \nu}} \ , &\nrmg {a}\cr 
\cc^{1,n}_{0,{\bar \nu}} ~=&~ \( \ \prod_{s=1}^{n-1}\ 
\( x_{i,\hl_{1,s}} + \nu_{i} - \hnu_{\hl_{1,s}} - s+1 \)\) 
\|\, \Psi_{0,{\bar \nu^{1,n}}}\, \|^2  \ ,  &\nrmg {b}\cr 
&\nu^{1,n}_{i\hj} ~=~ \nu^{1,n}_{k_1,\hj} ~=~ \nu^{1,n}_{i,\hl_{1,1}} 
~=~ \cdots ~=~ \nu^{1,n}_{i,\hl_{1,n}} 
~=~ 0 \ , \qquad \nu^{1,n}_{k_1,\hl_{1,n}} ~=~ 1 \ ,\cr 
&{\rm the\ rest\ of}~ \nu^{1,n}_{k,\hl}~ {\rm are\ as} ~\nu_{k,\hl} \cr 
\cc^{2,n}_{0,{\bar \nu}} ~=&~ \( x_{k_1,\hj} + \nu_{k_1} - \hnu_{\hj} \) 
\( \ \prod_{s=1}^{n-1}\ 
\( x_{i,\hl_{2,s}} + \nu_{i} - \hnu_{\hl_{2,s}} -s+1 \) \)
\times\cr &\times \|\, \Psi_{0,{\bar \nu^{2,n}}}\, \|^2 \ , &\nrmg {c}\cr 
&\nu^{2,n}_{i\hj} ~=~ \nu^{2,n}_{k_1,\hj} ~=~ \nu^{2,n}_{k_2,\hj} ~=~ 
\nu^{2,n}_{i,\hl_{2,1}} ~=~ \cdots ~=~ \nu^{2,n}_{i,\hl_{2,n}} 
~=~ 0 \ , \qquad \nu^{2,n}_{k_2,\hl_{2,n}} ~=~ 1 \ ,\cr 
&{\rm the\ rest\ of}~ \nu^{2,n}_{k,\hl}~ {\rm are\ as} ~\nu_{k,\hl} \cr 
\cc^{3,n}_{0,{\bar \nu}} ~=&~ \( x_{k_1,\hj} + \nu_{k_1} - \hnu_{\hj} \)\ 
\( x_{k_2,\hj} + \nu_{k_2} - \hnu_{\hj} +1 \)\ 
\times\cr &\times\ \( \ \prod_{s=1}^{n-1}\ 
\( x_{i,\hl_{3,s}} + \nu_{i} - \hnu_{\hl_{3,s}} -s+1 \) \)
\|\, \Psi_{0,{\bar \nu^{3,n}}}\, \|^2 \ , &\nrmg {d}\cr 
&\nu^{3,n}_{i\hj} ~=~ \nu^{3,n}_{k_1,\hj} ~=~ \nu^{3,n}_{k_2,\hj} ~=~ 
\nu^{3,n}_{k_3,\hj} ~=~ \nu^{3,n}_{i,\hl_{3,1}} 
~=~ \cdots ~=~ \nu^{3,n}_{i,\hl_{3,n}} 
~=~ 0 \ , \cr 
& \nu^{3,n}_{k_3,\hl_{3,n}} ~=~ 1 \cr 
&{\rm the\ rest\ of}~ \nu^{3,n}_{k,\hl}~ {\rm are\ as} ~\nu_{k,\hl} \cr 
}$$ 
{\it Proof:}~~ The reason for the counterterms is in the
transmutation of generators which happens for every pair from
\enux{}, \fatx{} by the following mechanism. Let us take one such
pair for fixed ~$(m,n)$. This means that ~$\Psi_{0,{\bar \nu}}$~ 
contains the operators: 
\eqna\intu
$$\eqalignno{\Psi_{0,{\bar \nu}} ~=& ~X^+_{i\hj}\, ...\, 
X^+_{k_m,\hj}\, ...\, X^+_{i,\hl_{m,n}}\, ...\, v_0 &\intu {a}}$$
and its norm squared is:
$$\eqalignno{ 
\|\, \Psi_{0,{\bar \nu}}\, \|^2 ~=&~ 
\( X^+_{i\hj}\, ...\, 
X^+_{k_m,\hj}\, ...\, X^+_{i,\hl_{m,n}}\, ...\, v_0\ , \ 
X^+_{i\hj}\, ...\, 
X^+_{k_m,\hj}\, ...\, X^+_{i,\hl_{m,n}}\, ...\, v_0 \) ~=\cr 
=&~ (-1)^\nu\ 
\(v_0\ ,\ ...\, X^-_{i,\hl_{m,n}}\, ...\, X^-_{k_m,\hj}\, ...\, 
X^-_{i\hj}\, X^+_{i\hj}\, ...\, X^+_{k_m,\hj}\, ...\,
X^+_{i,\hl_{m,n}}\, ...\, v_0\)\quad &\intu {b}}$$ 
Further we shall give only the term of ~$\|\, \Psi_{0,{\bar
\nu}}\, \|^2$~ which will turn into the discussed counterterm: 
$$\eqalignno{ 
\|\, \Psi_{0,{\bar \nu}}\, \|^2 ~\ap &~ (-1)^{\nu+1}\ 
\(v_0\ ,\ ...\, X^-_{i,\hl_{m,n}}\, ...\, X^-_{k_m,\hj}\, ...\, 
X^+_{i\hj}\, X^-_{i\hj}\, ...\, X^+_{k_m,\hj}\, ...\,
X^+_{i,\hl_{m,n}}\, ...\, v_0\) ~\ap \cr 
\ap&~ (-1)^{\nu+1}\ 
\(v_0\ ,\ ...\, X^-_{i,\hl_{m,n}}\, ...\, X^-_{k_m,\hj}\,
X^+_{i\hj}\, ...\, X^-_{i\hj}\, X^+_{k_m,\hj}\, ...\,
X^+_{i,\hl_{m,n}}\, ...\, v_0\) ~= \cr 
=&~ (-1)^{\nu+1}\ 
\Big(v_0\ ,\ ...\, X^-_{i,\hl_{m,n}}\, ...\, 
\big( - X^+_{i\hj}\, X^-_{k_m,\hj} - L^+_{i,k_m} \big)\, ...\, \times \cr 
&\times\, ...\, \big( - X^+_{k_m,\hj}\, X^-_{i\hj} - L^-_{k_m,i} \big)\, ...\,
X^+_{i,\hl_{m,n}}\, ...\, v_0\Big) ~\ap &\intu{c}\cr 
\ap&~ (-1)^{\nu+1}\ 
\(v_0\ ,\ ...\, X^-_{i,\hl_{m,n}}\, L^+_{i,k_m} \, ...\,
L^-_{k_m,i}\, X^+_{i,\hl_{m,n}}\, ...\, v_0\) ~= &\intu{d}\cr 
=&~ (-1)^{\nu+1}\ 
\Big(v_0\ ,\ ...\,\big( L^+_{i,k_m}\, X^-_{i,\hl_{m,n}} +
X^-_{k_m,\hl_{m,n}}\big)\, ...\, \times \cr 
&\times\, ...\, \big( X^+_{i,\hl_{m,n}}\, 
L^-_{k_m,i} + X^+_{k_m,\hl_{m,n}} \big)\,
 ...\, v_0\Big) ~\ap &\intu {e}\cr 
\ap&~ (-1)^{\nu+1}\ 
\(v_0\ ,\ ...\, X^-_{k_m,\hl_{m,n}}\, ...\, X^+_{k_m,\hl_{m,n}} 
 ...\, v_0\) ~= &\intu{f}\cr 
=&~-\ \|\, ...\, X^+_{k_m,\hl_{m,n}} 
 ...\, v_0\, \|^2  &\intu {g}\cr 
}$$ 
Thus, we have shown that the norm squared of ~$\Psi_{0,{\bar
\nu}}$~ contains a term which is the norm squared (with sign
'minus' - hence the word 'counterterm') of a state 
obtained from ~$\Psi_{0,{\bar\nu}}$~ by replacing the operators 
~$X^+_{i\hj}\,$, ~$X^+_{k_m,\hj}$~ and ~$X^+_{i,\hl_{m,n}}$~ by 
the operator ~$X^+_{k_m,\hl_{m,n}}\,$. Note that the latter was
not present in ~$\Psi_{0,{\bar\nu}}$~ due to the condition 
~$\nu_{k_m,\hl_{m,n}} ~=~ 0$~ in \fatx{}. Note also that the
counterterm state is of level ~$\nu-2$~ which brings the factor 
~$(-1)^{\nu-2}$~ in the passage from \intu{f} 
to \intu{g} which together with
the factor ~$(-1)^{\nu+1}$~ results in the overall minus sign in
\intu{g}. The described transmutation explains totally only
the first counterterm in \nrmg{b} obtained for ~$(m,n)=(1,1)$. 
The other counterterms get additional contributions, in
particular, from terms which we neglected in \intu{}. 
For the rest of the counterterms with ~$(m=1,n>1)$~
this affects the contributions of the operators 
~$X^+_{i,\hl_{1,s}}\,$, $s<n$. Analogously, for ~$m>1$~
this affects in addition the operators ~$X^+_{k_s,\hj}\,$,
$s<m$. In all cases, every counterterm is a polynomial in ~$d$~
of degree ~$\nu-2$. It remains only to explain the overall
restrictions on the number of of counterterms: since 
~$i<4,\ \hj>1$, it follows that ~$p ~\leq~ 4-i ~\leq ~3$, ~$q(m)
~\leq~ \hj-1$.\bsq

\nt 
Our next main result on the norms is:

\pr{6} Let ~$\Psi_{{\bar \ve}{\bar \nu}}$~ be a unfactorizable
state starting with the generator ~$Y^+_{i\hj}\,$. 
This means that there are one or more pairs
of integers ~$(k,\hl)$~ so that \facty{} is violated. 
Let us enumerate the pairs violating \facty{a} as:
\eqn\enuya{ (j_m,\hj_{m,n})\ , \quad i < j_1 < ... < j_p 
\ , \quad \hj < \hj_{m,1} < ... < \hj_{m,q(m)}} 
(note that ~$i<4,\ \hj<N$, ~$p ~\leq~ 4-i ~\leq ~3$, ~$q(m) ~\leq~ N-\hj$)
so that the following holds:
\eqn\fatya{ \ve_{j_m,\hj}~=~ \ve_{i,\hj_{m,n}} ~=~ 1 ~~{\rm and}~~
\ve_{j_m,\hj_{m,n}} ~=~ 0} 
Let us enumerate the pairs violating \facty{b} as:
\eqn\enuyb{ (k_m,\hk_{m,n})\ , \quad i<k_1 < ... < k_{p'} 
\ , \quad \hk_{m,1} > ... > \hk_{m,q'(m)}} 
(note that ~$i<4$, ~$p' ~\leq~ 4-i ~\leq ~3$, ~$q'(m) ~\leq~ N$)
so that the following holds:
\eqn\fatyb{ \ve_{k_m,\hj}~=~ \nu_{i,\hk_{m,n}} ~=~ 1 ~~{\rm and}~~
\nu_{k_m,\hk_{m,n}} ~=~ 0} 
Let us enumerate the pairs violating \facty{c} as:
\eqn\enuyc{ (\ell_m,\hl_{m,n})\ , \quad \ell_1 < ... < \ell_{p''} \
, \quad \hj< \hl_{m,1} < ... < \hl_{m,q''(m)}} 
(note that ~$\hj<N$, ~$p'' ~\leq~ 4$, ~$q''(m) ~\leq~ N-\hj$)
so that the following holds:
\eqn\fatyc{\ve_{i,\hl_{m,n}}~=~ \nu_{\ell_m,\hl_{m,n}} ~=~ 1
~~{\rm and}~~ \nu_{\ell_m,\hj} ~=~ 0} 
Then the norm is given by the following formula: 
\eqna\nrmu
$$\eqalignno{
\|\, \Psi_{{\bar \ve}{\bar \nu}}\, \|^2 ~~=
&~~ \(y_{i,\hj} +{\tilde \ve}_{i,\hj} + \nu_i + {\tnu}_\hj
\)\ \|\, \Psi_{{\bar \ve'}{\bar \nu}}\, \|^2 ~-~
\sum_{m=1}^p\ \sum_{n=1}^{q(m)}\ \cc^{m,n}_{{\bar \ve},{\bar
\nu}} \ ~-\cr &-~ \sum_{m=1}^{p'}\ \sum_{n=1}^{q'(m)}\
{\cc'}^{m,n}_{{\bar \ve},{\bar \nu}} \ ~-~ \sum_{m=1}^{p''}\
\sum_{n=1}^{q''(m)}\ {\cc''}^{m,n}_{{\bar \ve},{\bar \nu}} \
&\nrmu {a}\cr 
\cc^{1,n}_{{\bar \ve},{\bar \nu}} ~=&~ \( \ \prod_{s=1}^{n-1}\ 
\( y_{i,\hj_{1,s}} + \ve_i - \hve_{\hj_{1,s}} -s+1+ 
\nu_{i} + \hnu_{\hj_{1,s}}  \)\) 
\|\, \Psi_{{\bar \ve^{1,n}},{\bar \nu}}\, \|^2  \ ,  &\nrmu {b}\cr 
&\ve_i ~=~ \ve_{i,1} + \cdots + \ve_{i,N} \ , 
\quad \hve_\hl ~=~ \ve_{1,\hl} + \cdots + \ve_{4,\hl} \ , \cr 
&\ve^{1,n}_{i,\hj} ~=~ \ve^{1,n}_{j_1,\hj} ~=~ \ve^{1,n}_{i,\hj_{1,1}} 
~=~ \cdots ~=~ \ve^{1,n}_{i,\hj_{1,n}} 
~=~ 0 \ , \qquad \ve^{1,n}_{j_1,\hj_{1,n}} ~=~ 1 \ ,\cr 
&{\rm the\ rest\ of}~ \ve^{1,n}_{k,\hl}~ {\rm are\ as} ~\ve_{k,\hl} \cr 
\cc^{2,n}_{{\bar \ve},{\bar \nu}} ~=&~ 
\( y_{j_1,\hj} + \ve_{j_1} - \hve_{\hj} + \nu_{j_1} + \hnu_{\hj}  \)
\( \ \prod_{s=1}^{n-1}\ 
\( y_{i,\hj_{2,s}} + \ve_i - \hve_{\hj_{2,s}} -s+1+
\nu_{i} + \hnu_{\hj_{2,s}}  \)\) \times\cr & \times\ 
\|\, \Psi_{{\bar \ve^{2,n}},{\bar \nu}}\, \|^2  \ ,  &\nrmu {c}\cr 
&\ve^{2,n}_{i,\hj} ~=~ \ve^{2,n}_{j_1,\hj} ~=~ \ve^{2,n}_{j_2,\hj} ~=~ 
\ve^{2,n}_{i,\hj_{2,1}} ~=~ \cdots ~=~ \ve^{2,n}_{i,\hj_{2,n}} 
~=~ 0 \ , \qquad \ve^{2,n}_{j_2,\hj_{2,n}} ~=~ 1 \ ,\cr 
&{\rm the\ rest\ of}~ \ve^{2,n}_{k,\hl}~ {\rm are\ as} ~\ve_{k,\hl} \cr 
\cc^{3,n}_{{\bar \ve},{\bar \nu}} ~=&~ 
\( y_{j_1,\hj} + \ve_{j_1} - \hve_{\hj} + \nu_{j_1} + \hnu_{\hj}  \)
\( y_{j_2,\hj} + \ve_{j_2} - \hve_{\hj} +1+ \nu_{j_2} + \hnu_{\hj}  \)
\times\cr & \times \( \ \prod_{s=1}^{n-1}\ 
\( y_{i,\hj_{3,s}} + \ve_i - \hve_{\hj_{3,s}} - s+1+
\nu_{i} + \hnu_{\hj_{3,s}}  \)\) 
\|\, \Psi_{{\bar \ve^{3,n}},{\bar \nu}}\, \|^2  \ ,  &\nrmu {d}\cr 
&\ve^{3,n}_{i,\hj} ~=~ \ve^{3,n}_{j_1,\hj} ~=~ \ve^{3,n}_{j_2,\hj} ~=~ 
\ve^{3,n}_{j_3,\hj} ~=~ \ve^{3,n}_{i,\hj_{3,1}} ~=~ \cdots ~=~
\ve^{3,n}_{i,\hj_{3,n}} 
~=~ 0 \ , \qquad \ve^{3,n}_{j_3,\hj_{3,n}} ~=~ 1 \ ,\cr 
&{\rm the\ rest\ of}~ \ve^{3,n}_{k,\hl}~ {\rm are\ as}
~\ve_{k,\hl} \cr 
{\cc'}^{m,n}_{{\bar \ve},{\bar \nu}} ~=&~ 
\( \ \prod_{1\leq j \leq 4} \ \prod_{
{\hj\leq \hm \leq N} \atop {(j,\hm)\neq (i,\hj),
(k_m,\hj)}} \ 
\( y_{j\hm} + {\ve'}^m_j - \hve'_{\hm} + \nu'_j + {{\tilde {\nu'}}}^{m,n}_\hm
\)^{\ve_{j\hm}}\ \)\times \cr 
&\times \( \ \prod_{s=1}^{n-1}\ 
\( x_{i,\hk_{m,s}} + \nu_{i} - \hnu_{\hk_{m,s}} - s+1 \)\) 
\|\, \Psi_{0,{{\bar {\nu'}}}^{m,n}}\, \|^2  \ ,
&\nrmu {e}\cr  
&{\ve'}^m_j ~=~ \ve_{j,1} + \cdots + \ve_{j,N} - \d_{ji} - \d_{j,k_m} 
 \ , \quad \hve'_\hm ~=~ \ve_{1,\hm} + \cdots + \ve_{4,\hm} -
2\d_{\hm,\hj}\ , \cr 
&\nu'_j ~=~ \nu_j - \d_{ij} \ , \quad {\tilde {\nu'}}^{m,n}_\hm ~=~ 
{\tnu}_\hm - \d_{\hm,\hk_{m,n}} \ ,\cr 
&{\nu'}^{m,n}_{i,\hk_{m,1}} ~=~ \cdots ~=~ 
{\nu'}^{m,n}_{i,\hk_{m,n}} ~=~ 0 \ , \qquad
{\nu'}^{m,n}_{k_m,\hk_{m,n}} ~=~ 1 \ ,\cr 
&{\rm the\ rest\ of}~ {\nu'}^{m,n}_{k,\hl} ~ 
{\rm are\ as} ~ \nu_{k,\hl} \cr 
{\cc''}^{m,n}_{{\bar \ve},{\bar {\nu'}}} ~=&~ 
\( \ \prod_{1\leq j \leq 4} \ \prod_{
{\hj\leq \hm \leq N} \atop {(j,\hm)\neq (i,\hj),
(i,\hl_{m,n})}} \ 
\( y_{j\hm} + {\ve''}^m_j - \hve''_{\hm} + \nu''_j + {\tilde
{\nu''}}^{m,n}_\hm 
\)^{\ve_{j\hm}}\ \)\times \cr 
&\times \( \ \prod_{s=1}^{n-1}\ 
\( x_{\hl_m,\hl_{m,s}} + \nu_{\hl_m} - \hnu_{\hl_{m,s}} - s+1 \)\) 
\|\, \Psi_{0,{\bar {\nu''}}^{m,n}}\, \|^2  \ ,
&\nrmu {f}\cr  
&{\ve''}^m_j ~=~ \ve_{j,1} + \cdots + \ve_{j,N} - 2\d_{ji}
 \ , \quad \hve''_\hm ~=~ \ve_{1,\hm} + \cdots + \ve_{4,\hm} -
\d_{\hm,\hj}- \d_{\hm,\hl_{m,n}}\ , \cr 
&\nu''_j ~=~ \nu_j - \d_{j,\hl_m} \ , \quad {\tilde {\nu''}}^{m,n}_\hm ~=~ 
{\tnu}_\hm - \d_{\hm,\hl_{m,n}} \ ,\cr 
&{\nu''}^{m,n}_{\ell_m,\hl_{m,1}} ~=~ \cdots ~=~ 
{\nu''}^{m,n}_{\ell_m,\hl_{m,n}} ~=~ 0 \ , \qquad
{\nu''}^{m,n}_{\ell_m,\hj} ~=~ 1 \ ,\cr 
&{\rm the\ rest\ of}~ {\nu''}^{m,n}_{k,\hl} ~ 
{\rm are\ as} ~ \nu_{k,\hl} \ . \cr 
}$$ 
The ~{\it Proof}~ of this Proposition is analogous to the one of
Proposition 5, though more complicated since there are 
three possible mechanisms of transmutations corresponding to
the three exceptional situations given. Thus, in a case
described by \enuya{},\fatya{} the transmutation is:
\eqn\transa{
\Psi_{{\bar\ve},{\bar \nu}} ~= ~Y^+_{i\hj}\, ...\, 
Y^+_{j_m,\hj}\, ...\, Y^+_{i,\hj_{m,n}}\, ...\, v_0
~~\longrightarrow~~ ... \, Y^+_{j_m,\hj_{m,n}}\, ...\, v_0 }
In a case described by \enuyb{},\fatyb{} the transmutation is:
\eqn\transb{
\Psi_{{\bar\ve},{\bar \nu}} ~= ~Y^+_{i\hj}\, ...\, 
Y^+_{k_m,\hj}\, ...\, X^+_{i,\hk_{m,n}}\, ...\, v_0
~~\longrightarrow~~ ... \, X^+_{k_m,\hk_{m,n}}\, ...\, v_0 }
In a case described by \enuyc{},\fatyc{} the transmutation is:
\eqn\transc{
\Psi_{{\bar\ve},{\bar \nu}} ~= ~Y^+_{i\hj}\, ...\, 
Y^+_{i,\hl_{m,n}}\, ...\, X^+_{\ell_m,\hl_{m,n}}\, ...\, v_0
~~\longrightarrow~~ ... \, X^+_{\ell_m,\hj}\, ...\, v_0 }
Note that for ~$N=1$~ only the cases described by
\enuyb{},\fatyb{} are possible. Further we proceed as for
Proposition 5.\bsq 

\nt 
Our final main result on the norms is:

\pr{7}
If a state is not fully factorizable then the general expression
of its norm is: 
\eqna\nrmt
$$\eqalignno{
\|\, \Psi_{{\bar \ve}{\bar \nu}}\, \|^2 ~~=
&~~ \cn_{{\bar \ve}{\bar \nu}} ~-~ \cc_{{\bar \ve}{\bar \nu}} 
\ , &\nrmt {}}$$
where ~$\cc_{{\bar \ve}{\bar \nu}}$~ designates the possible
counterterms.\nl 
{\it Proof:}~~ This follows from Propositions 5 and 6. 
Consider first ~$\Psi_{0,{\bar \nu'}}$~ from Proposition 5. If it
is fully factorizable, then \nrmt{} follows at once. If it is not
fully factorizable but factorizable we first apply \ffx{} one or
more times until we are left with an unfactorizable state and 
then we apply Proposition 5 to the latter. We get another state
which plays the role of ~$\Psi_{0,{\bar \nu'}}$. Proceeding further
like this we establish \nrmt{} at the end. Analogously we
consider ~$\Psi_{{\bar\ve'},{\bar \nu}}$~ from Proposition 6 
until we establish \nrmt{} for this case.\bsq 

The above enables us to show that the conditions of the Theorem
are ~{\it sufficient}~ for ~$d~>~ d^-_{11}\,$. Indeed, in that
case ~$\cn_{{\bar \ve}{\bar \nu}} >0$~ for all states. 
What turns out to be important for the unitarity is that all 
counterterms are polynomials in ~$d$~ of lower degrees than 
~$\cn_{{\bar \ve}{\bar\nu}}$~ and all positivity requirements are
determined by the terms ~$\cn_{{\bar \ve}{\bar\nu}}$. Unitarity
at the reduction points will be considered in the next Section.\bsq

\newsec{Unitarity at the reduction points}

\subsec{The first reduction point}

\nt 
In this section we consider the unitarity of the irreps 
at the reducibility points ~$d^-_{i1}\,$. Unitarity is
established by noting that there are no negative norm states and
by factoring out the zero norm states which are a typical feature
of the Verma modules ~$V^\L$~ at the reducibility points. 
These zero norm states generate invariant submodules ~$I_{i1}$~ and are
decoupled in the factor modules ~$V^\L/I_{i1}$~ which realize the
UIRs at the points ~$d ~=~ d^-_{i1}\,$. 

In this subsection ~$d ~=~ d^-_{11}\,$, i.e., ~$x_{11} ~=~ 3$. 
We have the following:

\pr{8} Let ~$d ~=~ d^-_{11}\,$. There are no negative norm states. 
The zero norm states are described as follows. In the case
~$a_\hk ~\neq~ 0$, $\hk=1,...,N$, the states of zero norm ~$\cf^\L_0$~
from ~$\cf^\L$~ are given by ~$\Psi_{{\bar \ve}{\bar \nu}}$~ with: 
\eqn\frp{ \ve_{i\hj}= 0,1, \quad \nu_{i\hj} = \cases{
1 &~ $\hj=1$\cr 
0,1 &~otherwise}} 
The number of such states is ~$2^{8N-4}$~ and the number of oddly
generated states in the reduced irrep ~$L^\L ~\equiv \cf^\L/\cf^\L_0$~
is ~$15\times 2^{8N-4}$.\nl 
In the cases ~$a_1 ~=~ \cdots ~=~ a_\hk ~=~ 0$, $\hk ~=~
1,...,N-1$, in addition to those in \frp{} the following states
have zero norm: 
\eqna\frpa
$$\eqalignno{ &\ve_{i\hj}= 0,1, \cr 
& \nu_{i\hj} ~=~ \cases{
1 &~ $\hj=2$\cr 
0 &~ $i=\hj=1$\cr 
0,1 &~otherwise,} &\frpa{.1}\cr 
{\rm and}~~ &\nu_{i\hj} ~=~ \cases{1 &~ $\hj=3$\cr 
0 &~ $i=1, \hj=1,2$\cr 
0,1 &~otherwise,}&\frpa{.2}\cr 
& ... \cr
{\rm and}~~ &\nu_{i\hj} ~=~ \cases{1 &~ $\hj=\hk+1$\cr 
0 &~ $i=1, \hj=1,...,\hk $\cr 
0,1 &~otherwise,}&\frpa{.\tk}}$$ 
The number of states in \frpa{.1},\frpa{.2},...,\frpa{.\tk} is
~$2^{8N-5},2^{8N-6},...,2^{8N-4-\tk}$, resp., the overall number
of states in \frpa{} is ~$2^{8N-4-\tk}\(2^{\tk}-1\)$,
the number of states in the reduced ~$L^\L$~ - factoring out both
\frp{} and \frpa{} - is ~$2^{8N-4-\tk}\( 
2^{4+\tk} - 2^{\tk+1} +1\)$.\nl  
In the case ~$r_1 ~=~ 0$~ ($R$-symmetry scalars) in addition to
those in \frp{} and \frpa{} for ~$\hk=N-1$, the following states
have zero norm: 
\eqna\frpy
$$\eqalignno{ 
& \ve_{i\hj}= \cases{1 &~ $\hj=1$\cr 
0 &~ $i=1,\hj >1 $\cr 
0,1 &~otherwise,} \cr 
&\nu_{i\hj} ~=~ \cases{0 &~ $\hj=1$\cr 
0 &~ $i=1$\cr 
0,1 &~otherwise,}
&\frpy{.1}\cr 
{\rm and}~~ & \ve_{i\hj}= \cases{1 &~ $\hj=2$\cr 
0 &~ $i=1,\hj >2 $\cr 
0,1 &~otherwise,} \cr 
&\nu_{i\hj} ~=~ \cases{0 &~ $\hj=2$\cr 
0 &~ $i=1$\cr 
0,1 &~otherwise,}&\frpy{.2}\cr 
&...\cr
{\rm and}~~ & \ve_{i\hj}= \cases{1 &~ $\hj=N-1$\cr 
0 &~ $i=1,\hj =N$\cr 
0,1 &~otherwise,} \cr 
&\nu_{i\hj} ~=~ \cases{0 &~ $\hj=N-1$\cr 
0 &~ $i=1$\cr 
0,1 &~otherwise,} &\frpy{.N-1}\cr 
{\rm and}~~ & \ve_{i\hj}= \cases{1 &~ $\hj=N$\cr 
0,1 &~otherwise,} \cr 
&\nu_{i\hj} ~=~ \cases{0 &~ $\hj=N$\cr 
0 &~ $i=1$\cr 
0,1 &~otherwise,} &\frpy{.N}\cr 
}$$
The number of states in \frpy{.1},\frpy{.2},...,\frpy{.N-1},\frpy{.N} is
~$2^{6N-6},2^{6N-5},...,2^{7N-8}$,\nl $2^{7N-7}$, resp., the overall number
of states in \frpy{} is ~$2^{6N-6}\(2^{N}-1\)$,
the number of states in the reduced ~$L^\L$~ - factoring out 
\frp{}, \frpa{} (for $\tk=N-1$) and \frpy{} - is ~$2^{6N-6} \(
2^{2N+6} - (2^{N+3}+1) (2^{N}-1)\)$.\nl  
{\it Proof:}~~ There are no negative norm states if ~$d ~>~
d^-_{11}$~ and thus there are no such states for ~$d ~=~
d^-_{11}$~ by continuity. For the zero norm states we start with
the case ~$a_\hk ~\neq~ 0$, $\hk=1,...,N$. Inspecting formula
\nrmss{} we see that the fully factorized states of zero norm
have the form \frp{}. Indeed, the only factor in ~$\cn_{{\bar
\ve}{\bar \nu}}$~ that can be zero is ~$(x_{11} +{\tnu}_{11}) ~=~
(3+{\tnu}_{11})$, (hence ~$\nu_{11} ~=~ 1$), which happens if
~${\tnu}_{11} = -3$~ which happens if ~$\nu_{i1} ~=~ 1$,
$i=2,3,4$. (In general, ~$(x_{i\hj}+{\tnu}_{i\hj}) \geq
(3+{\tnu}_{i\hj}) \geq (i+\hj -2)$.) Further, the problem is
reduced to unfactorizable states. The main term of the norm
squared is given again by ~$\cn_{{\bar \ve}{\bar \nu}}$~ which
is zero. For further use we note more explicitly that for the states
from \frp{} we have: 
\eqn\fzz{ \cn_{{\bar \ve}{\bar \nu}} ~~\sim ~~
(x_{11}-3) (x_{21}-2) (x_{31}-1) x_{41} \ .} 
Now we shall show that also the counterterms are zero. For
this it is enough to show that ~$\nu^{m,n}_{i1} ~=~ 1$,
$i=1,2,3,4$~ in all auxiliary states that happen in the
counterterms. Consider first states starting with ~$X^+_{i\hj}$~
for which the norm is given in Proposition 6. The only way
~$\nu^{m,n}_{i1}$~ could differ from ~$\nu_{i1}$~ is if one of
the pairs in \enux{} is of the form $(k_m,1)$, more precisely,
that could be only one of the pairs ~$(k_m,\hk_{m,q(m)}) ~=~
(k_m,1)$. But then according to \fatx{} for any possible $m$ we
should have ~$\nu_{i,1} ~=~ 1$~ and ~$\nu_{k_m,1} ~=~ 0$~ which
does not hold. Thus, all counterterms are also zero. Consider
next states starting with ~$Y^+_{i\hj}$~ for which the norm is
given in Proposition 7. Here only the counterterms in \nrmu{e,f}
can possibly be non-zero. For the counterterm in \nrmu{e} the
only way ~${\nu'}^{m,n}_{i1}$~ could differ from ~$\nu_{i1}$~ is
if one of the pairs in \enuyb{} is of the form $(k_m,1)$, more
precisely, that could be only one of the pairs
~$(k_m,\tl_{m,q'(m)}) ~=~ (k_m,1)$. But then according to
\fatyb{} for any possible $m$ we should have ~$\nu_{i,1} ~=~ 1$~
and ~$\nu_{k_m,1} ~=~ 0$~ which does not hold. For the
counterterm in \nrmu{f} the considerations are simpler since it
is immediately seen from \enuyc{} that there is no pair that can
affect ~$\nu_{i,1}$~ since all ~$\tl_{m,n} > \hj \geq 1$, and if
we consider ~$\hj=1$~ then our state does not fulfil the
condition in \fatyc{} ~$\nu_{\ell_m,1} ~=~ 0$. Thus, all possible
counterterms are zero and thus all states in \frp{} have zero
norm. We continue with the cases ~$a_\hk ~\neq~ 0$,
$\hk=1,...,N-1$. Then ~$x_{1,\hk+1} ~=~ \cdots ~=~ x_{12} ~=~
x_{11} ~=~ 3$. Under the hypothesis in \frpa{.1} we have
~$\tnu_{12} ~=~ -3$, hence ~$x_{12} +\tnu_{12} ~=~ 0$~ and the
corresponding states have zero norm - the argument for
unfactorizable states goes analogously to above. The same
reasoning goes for all other cases in \frpa{}. 
For further use we note more explicitly that for the states
from \frpa{.\hl}, ~$\hl =1,2,...,\hk$, we have: 
\eqn\fza{ \cn_{{\bar \ve}{\bar \nu}} ~~\sim ~~ 
(x_{1,\hl+1}-3) (x_{2,\hl+1}-2) (x_{3,\hl+1}-1) x_{4,\hl+1} \ .} 
 We continue with
the case ~$r_1 ~=~ 0$. Then ~$y_{1,N} ~=~
\cdots ~=~ y_{11} ~=~ x_{11} ~=~ 3$.
Under the hypothesis in \frpy{.1} 
we have ~$\tve_{11} ~=~ -3$, hence ~$y_{11} +\tnu_{11} ~=~ 0$~
and the corresponding states have zero norm - the argument for
unfactorizable states goes analogously to above. The same
reasoning goes for all other cases in \frpy{}. 
For further use we note more explicitly that for the states
from \frpy{.\hl}, ~$\hl =1,2,...,N$, we have: 
\eqn\fzy{ \cn_{{\bar \ve}{\bar \nu}} ~~\sim ~~ 
 (y_{1,\hl}-3) (y_{2,\hl}-2) (y_{3,\hl}-1) y_{4,\hl} \ .}
The counting of states is straightforward.\bsq

\subsec{The other reduction points}

\nt 
We first consider the case \unt{b} of the Theorem: 
~$d ~=~ d^-_{21}$~ and ~$n_1 ~=~ 0$, i.e., ~$x_{11} ~=~ x_{21}
~=~ 2$. We have the following:

\pr{9} Let ~$d ~=~ d^-_{21}$~ and ~$n_1 ~=~ 0$. There are no
negative norm states. All states
of zero norm which are described in Proposition 8 have zero norm
also under the present hypothesis. There are further states of
zero norm which are described as follows. 
 In the case ~$a_\hk ~\neq~ 0$, $\hk=1,...,N$, the additional states
of zero norm are given by ~$\Psi_{{\bar \ve}{\bar \nu}}$~ with: 
\eqna\srp
$$\eqalignno{ 
&\ve_{i\hj}= 0,1, \cr 
& \nu_{11} + \nu_{21} +\nu_{31} +\nu_{41} = 3, \cr 
&\nu_{i\hj} = 0,1, \qquad \hj\neq 1,
&\srp{} }$$
The number of states in \srp{} is ~$2^{8N-2}$, and the number of
states in the reduced ~$L^\L$~ - factoring out both \frp{} and
\srp{} - is ~$11\times 2^{8N-4}$.\nl 
In the case ~$a_1 ~=~ 0$~ in addition to those in \srp{} the
following states have zero norm for ~$N=1$: 
\eqna\srpy
$$\eqalignno{ 
& \ve_{i1} = 1, \quad \nu_{11} =0, \quad \nu_{21} +\nu_{31} +
\nu_{41}  = 1, &\srpy{a} \cr  
{\rm and}~~ & \ve_{11} +\ve_{21} + \ve_{31} + \ve_{41} = 
3, \quad \nu_{i1} ~=~ 0 
&\srpy{b}\cr 
}$$
The number of states in \srpy{a}, \srpy{b}, is 3,4, resp. the
overall number of zero states - including \frp{}, \frpy{}, \srp{}, 
and \srpy{} - is 88, and thus the number of states
of the reduced ~$L^\L$~ is 168.\nl 
{\it Proof:}~~ We first have to show that the states of zero norm
from Proposition 8 have zero norm also here. With this we shall
establish also that there no negative norm states since those are
states are the only suspects for this. For the cases described by
\frp{} this follows by inspecting \fzz{} 
which is zero also here. In the cases described by
\frpa{\hl},\frpy{\hl} this follows by inspecting \fza{},\fzy{},
which are zero also here. Further, the proof is as of Proposition
8. In particular, for the states from \srp{} we have: 
\eqn\szz{ \cn_{{\bar \ve}{\bar \nu}} ~~\sim ~~
(x_{i_1,1}-2) (x_{i_2,1}-1) x_{i_3,1} \ ,} 
where ~$i_j$~ are from the set ~$1,2,3,4$, and thus at least one
of them is equal to ~$1$ or $2$, hence the RHS of \szz{} is zero.
Analogously, for the states from \srpy{a} holds \fzy{} for
~$\hl=1$, hence ~$\cn_{{\bar \ve}{\bar \nu}} =0$. 
For the states from \srpy{b} holds: 
\eqn\szy{ \cn_{{\bar \ve}{\bar \nu}} ~~\sim ~~
(y_{i_1,1}-2) (y_{i_2,1}-1) y_{i_3,1} \ ,} 
which is zero as \szz{} since ~$y_{i1} ~=~ x_{i1}$~ for
~$a_1=0$.\bsq

Next we consider the case \unt{c} of the Theorem: 
~$d ~=~ d^-_{31}$~ and ~$n_1 ~=~ n_2 ~=~ 0$, i.e., ~$x_{11} ~=~
x_{21}~=~ x_{31} ~=~ 1$. We have the following:

\pr{10} Let ~$d ~=~ d^-_{31}$~ and ~$n_1 ~=~ n_2 ~=~ 0$. 
There are no negative norm states. All states 
of zero norm which are described in Propositions 8 and 9 have zero norm
also under the present hypothesis. There are further states of
zero norm which are described as follows. 
 In the case ~$a_\hk ~\neq~ 0$, $\hk=1,...,N$, the additional states
of zero norm are given by ~$\Psi_{{\bar \ve}{\bar \nu}}$~ with: 
\eqn\trp{
\eqalign{ 
&\ve_{i\hj}= 0,1, \cr 
& \nu_{11}+ \nu_{21} +\nu_{31} +\nu_{41} = 2 }}
The number of states in \trp{} is ~$3\times 2^{8N-3}$, and the number of
states in the reduced ~$L^\L$~ - factoring out \frp{},
\srp{} and \trp{} - is ~$5 \times 2^{8N-4}$.\nl 
In the case ~$a_1 ~=~ 0$~ in addition to
those in \trp{} the following states have zero norm for ~$N=1$: 
\eqna\trpy
$$\eqalignno{ 
& \ve_{i1} = 1, \quad \nu_{i1} = \d_{i1}\ , &\trpy{a} \cr 
{\rm and}~~ & \ve_{11} + \ve_{21} + \ve_{31} + \ve_{41} = 3 ,
\quad \nu_{11} + \nu_{21} + \nu_{31} + \nu_{41} = 1, \cr 
&\ve_{11} = 1 ~\Rightarrow ~ \nu_{11} =0, \cr 
&\ve_{11} = 0 ~\Rightarrow ~ \nu_{21} =0, &\trpy{b} \cr 
{\rm and}~~ & \ve_{11} + \ve_{21} + \ve_{31} + \ve_{41} = 2 ,
 \quad \nu_{i1} = 0 &\trpy{c}}$$ 
The number of states in \trpy{a},\trpy{b},\trpy{c}, is 1,12,6, 
resp., the overall number of zero states - including \frp{},
\frpy{}, \srp{}, \srpy{}, \trp{}, \trpy{} - is 203, and thus 
the number of states of the reduced ~$L^\L$~ is 53.\nl
{\it Proof:}~~ 
We first have to show that the states of zero norm
from Proposition 8 and 9 have zero norm also here 
(establishing also the lack of negative norm states). For the
cases described by \frp{},\frpa{\hl},\frpy{\hl},\srp{},\srpy{} 
 this follows by inspecting
\fzz{},\fza{},\fzy{},\szz{},\szy{}, 
which are zero also here.
Further, the proof is as of Proposition 8 and 9. In particular,
for the states from \trp{} we have: 
\eqn\tzz{ \cn_{{\bar \ve}{\bar \nu}} ~~\sim ~~
(x_{i_1,1}-1) x_{i_2,1} \ ,} 
where ~$i_j$~ are from the set ~$1,2,3,4$, and thus at least one
of them is equal to ~$1$ or $2$ or $3$, hence the RHS of \tzz{} is zero.
For the states from \trpy{a}, resp., \trpy{b} hold \fzy{} for
~$\hl=1$, resp., \szy{}, hence ~$\cn_{{\bar \ve}{\bar \nu}} =0$. 
For the states from \trpy{c} we have: 
\eqn\tzy{ \cn_{{\bar \ve}{\bar \nu}} ~~\sim ~~
(y_{i_1,1}-1) y_{i_2,1} \ ,}
which is zero as \tzz{} since ~$y_{i1} ~=~ x_{i1}$~ for
~$a_1=0$.\bsq

Finally we consider the case \unt{d} of the Theorem:
~$d ~=~ d^-_{41}$~ and ~$n_1 ~=~ n_2 ~=~ n_3 ~=~ 0$, i.e.,
~$x_{11} ~=~ x_{21}~=~ x_{31} ~=~ x_{41} ~=~ 0$. We have the
following:

\pr{11} Let ~$d ~=~ d^-_{41}$~ and ~$n_1 ~=~ n_2 ~=~ n_3 ~=~ 0$.
There are no negative norm states. All states 
of zero norm which are described in Propositions
8,9 and 10 have zero norm also under the present hypothesis.
There are further states of zero norm which are described as
follows. In the case ~$a_\hk ~\neq~ 0$, $\hk=1,...,N$, the
additional states of zero norm are given by ~$\Psi_{{\bar
\ve}{\bar \nu}}$~ with: 
\eqn\qrp{
\eqalign{ 
&\ve_{i\hj}= 0,1, \cr 
& \nu_{11}+ \nu_{21} +\nu_{31} +\nu_{41} = 1, \cr 
&\nu_{i\hj} = 0,1, \qquad \hj\neq 1}}
The number of states in \qrp{} is ~$2^{8N-2}$, and the number of
states in the reduced ~$L^\L$~ - factoring out \frp{},
\srp{}, \trp{} and \qrp{} - is ~$2^{8N-4}$.\nl 
In the case ~$a_1 ~=~ 0$~ in addition to
those in \qrp{} the following states have zero norm for ~$N=1$: 
\eqna\qrpy
$$\eqalignno{ 
& \ve_{11} +\ve_{21} + \ve_{31} + \ve_{41} = 1 , \quad 
\nu_{i1} ~=~ 0 &\qrpy{} }$$
The number of states in \qrpy{} is 4, the overall number of zero
states - including \frp{}, \frpy{}, \srp{}, \srpy{}, \trp{}, \trpy{},
 \qrp{}, \qrpy{} - is ~$2^8-1$~ and thus 
the number of states of the reduced ~$L^\L$~ is 1, i.e., 
this is the trivial representation.\nl
{\it Proof:}~~ 
We first have to show that the states of zero norm
from Propositions 8,\ 9 and 10 have zero norm also here 
(establishing also the lack of negative norm states).
This is clear since in all cases the factor ~$\cn_{{\bar
\ve}{\bar \nu}}$~ contains as multiplicative factor some ~$x_{i1}$
and hence is zero. The same holds for the states from \qrp{}. 
For ~$N=1$~ and ~$a_1\neq 0$~ there are 16 states which are of
the form ~$\Psi_{{\bar\ve},0}$. For ~$a_1=0$~ all these, beside
the vacuum state, are of zero norm since the factor ~$\cn_{{\bar
\ve},0}$~ contains as multiplicative factor some ~$y_{i1}=x_{y1}=0$.
For the counting of states we have to note that the 16 states
in \srpy{a} and \trpy{a,b} are contained also in \qrp{} if ~$a_1
~=~ 0$.\bsq 

\newsec{Outlook} 

\nt 
In the last Subsection we gave the counting of states
in the cases ~$a_\hk ~\neq~ 0$, $\hk=1,...,N$, only for ~$N=1$. 
That would have taken many more pages due to the complicated
combinatorics for ~$N>1$~ when ~$a_\hk ~=~ 0$, and is left to a
subsequent publication. 

We also plan to construct the positive energy UIRs for ~$D=3,5$~
conformal supersymmetry taking up the corresponding conjectures
of Minwalla \Min. Other interesting objects are the conformal
superalgebras for ~$D>6$~ recently introduced in \DFLV. 

\bigskip 

\nt 
{\bf Acknowledgements.} ~~The author would like to thank S.
Ferrara for attracting his attention to the problem and for
discussions, and CERN-TH, where part of the work was done, for
hospitality. 

\bigskip 

\parskip=0pt \baselineskip=10pt

\listrefs

\np\end